\documentclass[10pt,american,prb,floatfix,superscriptaddress,twocolumn,showpacs,amsmath,amssymb,reprint]{revtex4-1}
\usepackage{ae,aecompl}
\usepackage[T1]{fontenc}
\usepackage[latin9]{inputenc}
\usepackage{color}
\usepackage{babel}
\usepackage{amsmath}
\usepackage{amssymb}
\usepackage{graphicx}
\usepackage[unicode=true,
 bookmarks=true,bookmarksnumbered=false,bookmarksopen=false,
 breaklinks=false,pdfborder={0 0 0},pdfborderstyle={},backref=false,colorlinks=true]
 {hyperref}
\hypersetup{
 pdfauthor={Dalson Almeida}}

\makeatletter
\usepackage{bbm}

\makeatother

\begin{document}

\title{Induced spin-triplet pairing in the coexistence state of antiferromagnetism
and singlet superconductivity: collective modes and microscopic properties}

\author{D. E. Almeida}

\affiliation{Instituto de Física Gleb Wataghin, Unicamp, Rua Sérgio Buarque de
Holanda, 777, CEP 13083-859 Campinas, SP, Brazil}

\affiliation{School of Physics and Astronomy, University of Minnesota, Minneapolis
55455, USA}

\author{R. M. Fernandes}

\affiliation{School of Physics and Astronomy, University of Minnesota, Minneapolis
55455, USA}

\author{E. Miranda}

\affiliation{Instituto de Física Gleb Wataghin, Unicamp, Rua Sérgio Buarque de
Holanda, 777, CEP 13083-859 Campinas, SP, Brazil}
\begin{abstract}
The close interplay between superconductivity and antiferromagnetism
in several quantum materials can lead to the appearance of an unusual
thermodynamic state in which both orders coexist microscopically,
despite their competing nature. A hallmark of this coexistence state
is the emergence of a spin-triplet superconducting gap component,
called $\pi$-triplet, which is spatially modulated by the antiferromagnetic
wave-vector, reminiscent of a pair-density wave. In this paper, we
investigate the impact of these $\pi$-triplet degrees of freedom
on the phase diagram of a system with competing antiferromagnetic
and superconducting orders. Although we focus on a microscopic two-band
model that has been widely employed in studies of iron pnictides,
most of our results follow from a Ginzburg-Landau analysis, and as
such should be applicable to other systems of interest, such as cuprates
and heavy fermions. The Ginzburg-Landau functional reveals not only
that the $\pi$-triplet gap amplitude couples tri-linearly with the
singlet gap amplitude and the staggered magnetization magnitude, but
also that the $\pi$-triplet $d$-vector couples linearly with the
magnetization direction. While in the mean field level this coupling
forces the $d$-vector to align parallel or anti-parallel to the magnetization,
in the fluctuation regime it promotes two additional collective modes
\textendash{} a Goldstone mode related to the precession of the $d$-vector
around the magnetization and a massive mode, related to the relative
angle between the two vectors, which is nearly degenerate with a Leggett-like
mode associated with the phase difference between the singlet and
triplet gaps. We also investigate the impact of magnetic fluctuations
on the superconducting-antiferromagnetic phase diagram, showing that
due to their coupling with the $\pi$-triplet order parameter, the
coexistence region is enhanced. This effect stems from the fact that
the $\pi$-triplet degrees of freedom promote an effective attraction
between the antiferromagnetic and singlet superconducting degrees
of freedom, highlighting the complex interplay between these two orders,
which goes beyond mere competition for the same electronic states.
\end{abstract}
\maketitle

\section{Introduction}

\label{sec:intro}

The close proximity between the superconducting (SC) and antiferromagnetic
(AFM) transitions in unconventional superconductors such as cuprates,
iron pnictides, and heavy fermions, has motivated a profound investigation
of the interplay between these two phases \cite{Sachdev03,SCZhang04,Scalapino12,Fradkin15}.
In general, these two ordered states compete for the same electronic
states, as manifested for instance by the suppression of the AFM order
parameter below the SC transition temperature $T_{\mathrm{c}}$ observed
in neutron diffraction experiments \cite{Fe_coexist_1,HF_coexist_1}.
Despite this competition, these two antagonistic phases can coexist
microscopically, giving rise to a new thermodynamic state in which
both the $\mathrm{U}(1)$ gauge symmetry and the $\mathrm{SO}(3)$
spin-rotational symmetry are simultaneously broken. Experimentally,
identifying such a microscopic coexistence phase is challenging: because
bulk probes are generally sensitive not only to the order parameter,
but also to its volume fraction \cite{Fe_coexist_1}, it is difficult
to distinguish the situation in which the two orders coexist locally
from the case in which the system phase-separates into non-overlapping
domains of AFM and SC orders. As a result, local probes are generally
needed to unambiguously identify the AFM-SC microscopic coexistence
phase.

Recently, NMR (nuclear magnetic resonance), $\mu$SR (muon spin rotation),
STM (scanning tunneling microscopy), and Mössbauer experiments have
revealed that several iron-based superconductors display this unique
AFM-SC coexistence state in their phase diagrams \cite{Fe_coexist_2,Fe_coexist_3,Fe_coexist_4,Fe_coexist_6,Fe_coexist_7,Fe_coexist_8,Fe_coexist_9}.
Data supporting the existence of this state in certain cuprates \cite{cuprates_coexist}
and heavy fermions \cite{HF_coexist_1,HF_coexist_2,HF_coexist_3,Rosa17}
have also been reported. Thus, it is of general interest to elucidate
the microscopic and macroscopic properties of the AFM-SC coexistence
state, not only to provide useful benchmarks to probe it, but also
to search for possible novel phenomena in this unusual phase of matter. 

Indeed, many theoretical works have tackled this issue and provided
invaluable information about the interplay between AFM and unconventional
SC in the coexistence state \cite{Baltensperger63,Bulaevskii80,Grest81,Machida81,Gabovich84,Gulacsi86,Machida88,Sachdev02,Tesanovic08,Das08,Chubukov09,Fernandes10,Ghaemi11,Timm14,DHLee14,Eremin15,Eremin16}.
Interestingly, as shown in Ref. \cite{triplet_organics}, the fact
that the AFM order parameter $\boldsymbol{M}$ and the singlet SC
order parameter $\Delta_{\mathrm{s}}$ are simultaneously non-zero
implies that a triplet component of the superconducting order parameter
is generated, $\Delta_{\mathrm{t}}\propto\Delta_{\mathrm{s}}M$. It
is clear that this triplet component only exists in the case of microscopic
AFM-SC phase coexistence, since in the case of phase separation, either
$\Delta_{\mathrm{s}}$ or $\boldsymbol{M}$ vanish at an arbitrary
point. Consequently, detecting this triplet component, often called
$\pi$-triplet (and hereafter denoted t-SC), would provide unambiguous
evidence in favor of a coexistence AFM-SC state. On the microscopic
level, this triplet component pairs electrons with momenta $-\boldsymbol{k}$
and $\boldsymbol{k}+\boldsymbol{Q}$, i.e.

\begin{equation}
\Delta_{\mathrm{t}}\propto\sum_{\boldsymbol{k}}\left(\hat{\boldsymbol{d}}\cdot\boldsymbol{\sigma}\mathrm{i}\sigma^{y}\right)_{ss^{\prime}}^{\dagger}\langle c_{\boldsymbol{k}+\boldsymbol{Q},s}c_{-\boldsymbol{k},s^{\prime}}\rangle\mathrm{.}\label{main_eq}
\end{equation}
Here, $c_{\boldsymbol{k},s}$ is the standard fermionic operator associated
with an electron with momentum $\boldsymbol{k}$ and spin $s$, $\sigma^{j}$
are the Pauli matrices, and $\hat{\boldsymbol{d}}$ is the triplet
$d$-vector. Since the center-of-mass of the Cooper pair has momentum
$\boldsymbol{Q}$ equal to the AFM wave-vector, this order parameter
behaves similarly to a pair density-wave \cite{PDW1,PDW2,PDW3}. Note
however that the term pair-density wave has been primarily employed
to describe a SC state without a homogeneous gap component, which
is not possible in our case, since $\Delta_{\mathrm{t}}$ only appears
in the presence of a homogeneous singlet component $\Delta_{\mathrm{s}}$.
Despite the fact that the system still has inversion symmetry and
$\Delta_{\mathrm{t}}$ has even parity, we identify $\Delta_{\mathrm{t}}$
as a triplet because of its spin structure. The reason why the triplet
spin structure is allowed in the AFM phase, even though $\Delta_{\mathrm{t}}$
has even parity, is because inside the AFM phase $c_{\boldsymbol{k}+\boldsymbol{Q},s}$
and $c_{\boldsymbol{k},s}$ become different ``electronic flavors''
due to the band folding. This can be more easily visualized by rewritting
the expression for the triplet gap as:

\begin{equation}
\Delta_{\mathrm{t}}\propto\frac{1}{2}\sum_{\boldsymbol{k}}\left(\hat{\boldsymbol{d}}\cdot\boldsymbol{\sigma}\mathrm{i}\sigma^{y}\right)_{ss^{\prime}}^{\dagger}\left(i\tau^{y}\right)^{\mu\nu}\left\langle \Phi_{\mathbf{k},s}^{\mu}\Phi_{-\mathbf{k},s}^{\mu}\right\rangle 
\end{equation}
where $\Phi_{\mathbf{k},s}^{\mu}=\left(\begin{array}{cc}
c_{\boldsymbol{k}+\boldsymbol{Q},s} & c_{-\boldsymbol{k},s}\end{array}\right)^{T}$ is a spinor in both spin space and AFM band-folded space. The situation
resembles the case of multi-orbital systems with atomic spin-orbit
coupling $\boldsymbol{S}\cdot\boldsymbol{L}$, in which case the superconducting
order parameter generally has both singlet and triplet components
(although inversion symmetry is preserved) \cite{Vafek13}.

Different aspects of the impact of this $\pi$-triplet component on
the AFM-SC coexistence phase have been previously discussed \cite{triplet_organics,triplet_Fukuyama,triplet_Kyung,triplet_heavy_fermions,triplet_heavy_fermions2,triplet_heavy_fermions3,triplet_GL,triplet_Fe_pnictides}.
In most cases, the analyses focused on the ordered state, where the
$d$-vector is fixed parallel to the magnetization direction $\hat{\boldsymbol{M}}$.
In this work, we focus instead on the disordered state, and investigate
the coupling between the $d$-vector and the magnetization $\boldsymbol{M}$.
For concreteness, we consider a microscopic two-band model widely
employed to study the interplay between AFM and SC in the iron pnictide
superconductors, but most of our results should hold in other systems
as well. As it was previously shown in Refs. \cite{Chubukov09,Fernandes10},
the phase diagram of this model displays a tetracritical point and,
consequently, an AFM-SC coexistence phase. Near the tetracritical
point, we then use the microscopic model to derive the Ginzburg-Landau
free energy in the disordered state and show that the $d$-vector
couples linearly with $\boldsymbol{M}$. While in the ordered state
this implies that the two vectors are parallel, as assumed in previous
works, in the disordered state it gives rise to a collective mode
corresponding to oscillations of the angle between the AFM order parameter
and the $d$-vector of the t-SC order parameter. We find that in general
this collective mode has a finite energy, which is comparable to,
but larger than, the Leggett-like mode associated with oscillations
of the relative phase between the singlet and triplet SC order parameters.

We then go beyond the mean field approach and study how magnetic fluctuations
modify the phase diagram. In general, we find that AFM fluctuations
shrink the magnetically ordered region, as expected, while keeping
the second-order character of the phase transition lines. More importantly,
the t-SC order acts to expand the AFM-SC phase coexistence, by promoting
an effective attraction between these two otherwise competing orders.
Finally, we discuss the implications of our results to the understanding
of the phase diagrams of unconventional superconductors.

The paper is organized as follows: in Section II we present our microscopic
model and derive the Ginzburg-Landau functional. The mean field phase
diagram and the analysis of the corresponding collective modes are
shown in Section III. Section IV is devoted to the investigation of
the effects of magnetic fluctuations. Concluding remarks are presented
in Section V. Two Appendices contain additional technical details
of the derivations discussed in the main text. 

\section{Microscopic model and Ginzburg-Landau functional}

\label{sec:modelMFGL}

\subsection{The model}

We consider a two-dimensional two-band model containing one hole band
and one electron band. Such a model has been widely employed in studies
of the interplay between SC and AFM in iron pnictides, see for instance
Refs. \cite{Chubukov09,Fernandes10}. While this model is useful to
obtain microscopic values for the Ginzburg-Landau parameters, we emphasize
that most of our results are quite general and apply to any other
system where the AFM and SC transition lines meet at a tetracritical
point. The Hamiltonian contains four terms 

\begin{equation}
\mathcal{H}=\mathcal{H}_{0}+\mathcal{H}_{\mathrm{AFM}}+\mathcal{H}_{\mathrm{SC}}^{\mathrm{s}}+\mathcal{H}_{\mathrm{SC}}^{\mathrm{t}}.\label{eq:H}
\end{equation}

The noninteracting part $\mathcal{H}_{0}$ describes the two bands,
whose centers are displaced by $\boldsymbol{Q}=\left(\pi,\pi\right)$
\begin{equation}
\mathcal{H}_{0}=\sum_{\boldsymbol{k},s}\left(\xi_{1,\boldsymbol{k}}c_{\boldsymbol{k},s}^{\dagger}c_{\boldsymbol{k},s}+\xi_{2,\boldsymbol{k}+\boldsymbol{Q}}f_{\boldsymbol{k}+\boldsymbol{Q},s}^{\dagger}f_{\boldsymbol{k}+\boldsymbol{Q},s}\right),\label{eq:H_0}
\end{equation}
where $c_{\boldsymbol{k},s}^{\dagger}$ ($f_{\boldsymbol{k}+\boldsymbol{Q},s}^{\dagger}$)
is an operator that creates a fermion with momentum $\boldsymbol{k}$
($\boldsymbol{k}+\boldsymbol{Q}$) and spin projection $s=\pm1$.
The isotropic hole-band dispersion is given by $\xi_{1,\boldsymbol{k}}=\varepsilon_{1,0}-k^{2}/2m-\mu$,
whereas the anisotropic electron-band dispersion is $\xi_{2,\boldsymbol{k}+\boldsymbol{Q}}=-\varepsilon_{2,0}+k_{x}^{2}/2m_{x}+k_{y}^{2}/2m_{y}-\mu$.
Note that the chemical potential $\mu$ has been included in the dispersions
and $\varepsilon_{1,0}>0$ and $\varepsilon_{2,0}>0$ are offset energies.
To proceed, we introduce the notation $\tan\theta=k_{y}/k_{x}$ and
rewrite the band dispersions according to $\xi_{2,\boldsymbol{k}+\boldsymbol{Q}}=-\xi_{1,\boldsymbol{k}}+2\delta_{\boldsymbol{k}}$,
where $\delta_{\boldsymbol{k}}=\delta_{0}(k)+\delta_{2}(k)\cos2\theta$
measures the deviation from the perfect nesting condition ($\xi_{1,\boldsymbol{k}}=-\xi_{2,\boldsymbol{k+Q}}$)
with $\delta_{0}(k)=(\varepsilon_{1,0}-\varepsilon_{2,0})/2-\mu+k^{2}(m_{x}^{-1}+m_{y}^{-1}-2m^{-1})/8$
and $\delta_{2}(k)=k^{2}(m_{x}^{-1}-m_{y}^{-1})/8$ \cite{Chubukov09}.

The second term of the Hamiltonian describes the repulsive interactions
that drive AFM
\begin{equation}
\mathcal{H}_{\mathrm{AFM}}=-\frac{V_{\mathrm{m}}}{2\upsilon}\sum_{\boldsymbol{k},\boldsymbol{k}^{\prime}}\left(c_{\boldsymbol{k},s}^{\dagger}\boldsymbol{\sigma}_{ss^{\prime}}f_{\boldsymbol{k}+\boldsymbol{Q},s^{\prime}}\right)\cdot\left(f_{\boldsymbol{k}^{\prime}+\boldsymbol{Q},\sigma}^{\dagger}\boldsymbol{\sigma}_{\sigma\sigma^{\prime}}c_{\boldsymbol{k}^{\prime},\sigma^{\prime}}\right)\mathrm{,}\label{H_AFM}
\end{equation}
where $\upsilon$ is the volume of the system, $V_{\mathrm{m}}$ is
the coupling constant (whose momentum dependence we dropped, for simplicity),
$\boldsymbol{\sigma}_{ss^{\prime}}$ is the ($ss^{\prime}$) element
of the Pauli matrix vector. Hereafter, repeated spin indices are implicitly
summed over.

The fermions are also subject to inter-band pairing interactions,
both in the singlet and in the triplet channels, which are described,
respectively, by the two last terms in $\mathcal{H}$
\begin{eqnarray}
\mathcal{H}_{\mathrm{SC}}^{\mathrm{s}} & = & -\frac{V_{\mathrm{s}}}{2\upsilon}\sum_{\boldsymbol{k},\boldsymbol{k}^{\prime}}\left[\left(\mathrm{i}\sigma^{y}\right)_{ss^{\prime}}\left(\mathrm{i}\sigma^{y}\right){}_{\sigma\sigma^{\prime}}^{\dagger}\right.\nonumber \\
 &  & \left.\left(c_{\boldsymbol{k},s}^{\dagger}c_{-\boldsymbol{k},s^{\prime}}^{\dagger}f_{-\boldsymbol{k}^{\prime}-\boldsymbol{Q},\sigma}f_{\boldsymbol{k}^{\prime}+\boldsymbol{Q},\sigma^{\prime}}\right)+\mathrm{h.c.}\right]\label{eq:H_SSC}
\end{eqnarray}
and
\begin{eqnarray}
\mathcal{H}_{\mathrm{SC}}^{\mathrm{t}} & = & -\frac{V_{\mathrm{t}}}{2\upsilon}\sum_{\boldsymbol{k},\boldsymbol{k}^{\prime}}\left[\left(\hat{\boldsymbol{d}}\cdot\boldsymbol{\sigma}\mathrm{i}\sigma^{y}\right)_{ss^{\prime}}\left(\hat{\boldsymbol{d}}\cdot\boldsymbol{\sigma}\mathrm{i}\sigma^{y}\right)_{\sigma\sigma^{\prime}}^{\dagger}\right.\nonumber \\
 &  & \left.\left(f_{\boldsymbol{k}+\boldsymbol{Q},s}^{\dagger}c_{-\boldsymbol{k},s^{\prime}}^{\dagger}c_{-\boldsymbol{k}^{\prime},\sigma}f_{\boldsymbol{k}^{\prime}+\boldsymbol{Q},\sigma^{\prime}}\right)+\mathrm{h.c.}\right]\mathrm{,}\label{eq:H_TSC}
\end{eqnarray}
where $V_{\mathrm{s}}$ and $V_{\mathrm{t}}$ are the singlet and
triplet SC couplings, respectively. The triplet SC pairs have a finite
momentum $\boldsymbol{Q}$, and are characterized by the unitary $d$-vector
$\hat{\boldsymbol{d}}=(\hat{d}_{x},\hat{d}_{y},\hat{d}_{z})^{T}$.
The spinor $\hat{\boldsymbol{d}}\cdot\boldsymbol{\sigma}\mathrm{i}\sigma^{y}$
for triplet Cooper pairs is discussed in Refs. \cite{Balian63,Leggett75,Maeno03}
(see also Refs. \cite{triplet_organics,triplet_heavy_fermions}).
We follow Ref. \cite{triplet_Kyung} and include from the beginning
the triplet component because, when the rotational symmetry in spin
space is broken and the system undergoes a singlet SC phase transition,
triplet components $\langle c_{-\boldsymbol{k},s}f_{\boldsymbol{k}+\boldsymbol{Q},s^{\prime}}\rangle$
are necessarily generated.

We now define the various order parameters (OP). The staggered magnetization
is
\begin{equation}
\boldsymbol{M}=-\frac{V_{\mathrm{m}}}{2\upsilon}\sum_{\boldsymbol{\boldsymbol{k}}}\boldsymbol{\sigma}_{ss^{\prime}}\langle f_{\boldsymbol{k}+\boldsymbol{Q},s}^{\dagger}c_{\boldsymbol{k},s^{\prime}}\rangle\mathrm{,}\label{OP_M}
\end{equation}
the singlet SC OPs for each band are
\begin{eqnarray}
\Delta_{\mathrm{s},1} & = & -\frac{V_{\mathrm{s}}}{\upsilon}\sum_{\boldsymbol{k}}\langle f_{\boldsymbol{k}+\boldsymbol{Q},\uparrow}f_{-\boldsymbol{k}-\boldsymbol{Q},\downarrow}\rangle\mathrm{,}\label{OP_singlet1}\\
\Delta_{\mathrm{s},2} & = & -\frac{V_{\mathrm{s}}}{\upsilon}\sum_{\boldsymbol{k}}\langle c_{\boldsymbol{k},\uparrow}c_{-\boldsymbol{k},\downarrow}\rangle,\label{OP_singlet2}
\end{eqnarray}
and the triplet SC OP is \cite{Balian63,Leggett75,triplet_organics,triplet_heavy_fermions}
\begin{equation}
\Delta_{\mathrm{t}}=-\frac{V_{\mathrm{t}}}{2\upsilon}\sum_{\boldsymbol{k}}\langle f_{\boldsymbol{k}+\boldsymbol{Q},s}\left(\hat{\boldsymbol{d}}\cdot\boldsymbol{\sigma}\mathrm{i}\sigma^{y}\right)_{ss^{\prime}}^{\dagger}c_{-\boldsymbol{k},s^{\prime}}\rangle\mathrm{.}\label{OP_triplet}
\end{equation}

We use the usual mean field decoupling to rewrite the quartic terms
of $\mathcal{H}$ as
\begin{equation}
\mathcal{H}_{\mathrm{AFM}}\approx\sum_{\boldsymbol{k}}\left[c_{\boldsymbol{k},s}^{\dagger}\left(\boldsymbol{M}\cdot\boldsymbol{\sigma}\right){}_{ss^{\prime}}f_{\boldsymbol{k}+\boldsymbol{Q},s^{\prime}}+\mathrm{h.c.}\right]
\end{equation}
\begin{equation}
\mathcal{H}_{\mathrm{SC}}^{\mathrm{s}}\approx\sum_{\boldsymbol{k}}\left(\Delta_{\mathrm{s},1}c_{\boldsymbol{k},\uparrow}^{\dagger}c_{-\boldsymbol{k},\downarrow}^{\dagger}+\Delta_{\mathrm{s},2}f_{\boldsymbol{k}+\boldsymbol{Q},\uparrow}^{\dagger}f_{-\boldsymbol{k}-\boldsymbol{Q},\downarrow}^{\dagger}+\mathrm{h.c.}\right)
\end{equation}
 and
\begin{align}
\mathcal{H}_{\mathrm{SC}}^{\mathrm{t}} & \approx-\frac{1}{2}\sum_{\boldsymbol{k}}\left[\left(\hat{\Delta}_{\mathrm{t}}\right)_{ss'}\left(f_{\boldsymbol{k}+\boldsymbol{Q},s}^{\dagger}c_{-\boldsymbol{k},s^{\prime}}^{\dagger}-c_{\boldsymbol{k},s}^{\dagger}f_{-\boldsymbol{k}-\boldsymbol{Q},s'}^{\dagger}\right)\right.\nonumber \\
 & \left.+\mathrm{h.c.}\right]\mathrm{,}
\end{align}
where we introduced the notation $\left(\hat{\Delta}_{\mathrm{t}}\right)_{ss'}=\left(\hat{\boldsymbol{d}}\cdot\boldsymbol{\sigma}\mathrm{i}\sigma^{y}\right)_{ss^{\prime}}\Delta_{\mathrm{t}}$
and we also omitted the constant terms for simplicity. Note that the
singlet SC gap of one band is due to the action of the electrons in
the other band and that the triplet SC OP has a finite momentum $\boldsymbol{Q}$.
To proceed, we introduce the eight-component Balian-Werthamer spinor
$\psi_{\boldsymbol{k}}^{\dagger}=(c_{\boldsymbol{k},\uparrow}^{\dagger},c_{\boldsymbol{k},\downarrow}^{\dagger},c_{-\boldsymbol{k},\uparrow},c_{-\boldsymbol{k},\downarrow},f_{\boldsymbol{k}+\boldsymbol{Q},\uparrow}^{\dagger},f_{\boldsymbol{k}+\boldsymbol{Q},\downarrow}^{\dagger},f_{-\boldsymbol{k}-\boldsymbol{Q},\uparrow},f_{-\boldsymbol{k}-\boldsymbol{Q},\downarrow}$)
to write the total Hamiltonian in compact form as
\begin{equation}
\mathcal{H}_{\mathrm{MF}}=\frac{1}{2}\sum_{\boldsymbol{k}}\psi_{\boldsymbol{k}}^{\dagger}\hat{H}_{\boldsymbol{k}}\psi_{\boldsymbol{k}}+E_{\mathrm{cond}}\mathrm{,}\label{H_MF}
\end{equation}
where $E_{\mathrm{cond}}=2\upsilon\left[\frac{\boldsymbol{M}^{2}}{V_{\mathrm{m}}}-\frac{\mathrm{Re}(\Delta_{\mathrm{s},1}\Delta_{\mathrm{s},2}^{*})}{V_{\mathrm{s}}}+\frac{|\Delta_{\mathrm{t}}|^{2}}{V_{\mathrm{t}}}\right]$
contains the constant terms omitted above and
\begin{equation}
\hat{H}_{\boldsymbol{k}}=\begin{bmatrix}\xi_{1,\boldsymbol{k}}\mathbbm{1}_{2} & \Delta_{\mathrm{s},1}\left(i\sigma^{y}\right) & \boldsymbol{M}\cdot\boldsymbol{\sigma} & \hat{\Delta}_{\mathrm{t}}\\
-\Delta_{\mathrm{s},1}^{*}\left(i\sigma^{y}\right) & -\xi_{1,\boldsymbol{k}}\mathbbm{1}_{2} & -\hat{\Delta}_{\mathrm{t}}^{\dagger} & -\left(\boldsymbol{M}\cdot\boldsymbol{\sigma}\right)\\
\left(\boldsymbol{M}\cdot\boldsymbol{\sigma}\right)^{T} & -\hat{\Delta}_{\mathrm{t}} & \xi_{2,\boldsymbol{k}+\boldsymbol{Q}}\mathbbm{1}_{2} & \Delta_{\mathrm{s},2}\left(i\sigma^{y}\right)\\
\hat{\Delta}_{\mathrm{t}}^{\dagger} & -\left(\boldsymbol{M}\cdot\boldsymbol{\sigma}\right)^{T} & -\Delta_{\mathrm{s},2}^{*}\left(i\sigma^{y}\right) & -\xi_{2,\boldsymbol{k}+\boldsymbol{Q}}\mathbbm{1}_{2}
\end{bmatrix}\mathrm{.}\label{reduced-H}
\end{equation}
Note that we have omitted the constant term $\frac{1}{2}\sum_{\boldsymbol{k}}\left(\xi_{1,\boldsymbol{k}}+\xi_{2,\boldsymbol{k}+\boldsymbol{Q}}\right)$
in Eq. (\ref{H_MF}).

Because $\boldsymbol{Q}$ is commensurate and $2\boldsymbol{Q}$ is
a reciprocal lattice vector, the magnetic OP $\boldsymbol{M}$ is
real. Furthermore, we assume that $V_{s}>0$, implying that the SC
gaps are of equal magnitude but different signs in the two bands,
$\Delta_{\mathrm{s},2}=-\Delta_{\mathrm{s},1}=\Delta_{\mathrm{s}}$,
as discussed in Ref. \cite{Fernandes10}. This is the so-called $s^{+-}$
superconducting state. As usual, the gaps are parametrized by their
magnitude and phases, $\Delta_{\mathrm{s}}=|\Delta_{\mathrm{s}}|e^{\mathrm{i}\alpha_{\mathrm{s}}}$
and $\Delta_{\mathrm{t}}=|\Delta_{\mathrm{t}}|e^{\mathrm{i}(\alpha_{\mathrm{s}}-\alpha_{\mathrm{st}})}$.
Note that, in the present analysis, we will ignore modes associated
with the relative phase between the two gaps of the two bands \textendash{}
such modes usually have high energies when the pairing interaction
is dominated by inter-band processes, as in our case \cite{Benfatto13}.
Furthermore, they are absent in single band systems with a $d$-wave
gap, for which the present analysis can be extended in a straightforward
way.

\subsection{Derivation of the free energy}

The model discussed above was previously shown to display an AFM-SC
tetracritical point, for a wide range of band dispersion parameters
\cite{Chubukov09,Fernandes10} (note that in the case of conventional
$s^{++}$ SC, the phase diagram has only a bicritical point and no
AFM-SC coexistence). The Ginzburg-Landau free energy can be obtained
in a straightforward way by integrating out the fermionic fields of
the quadratic mean field Hamiltonian {[}Eq. (\ref{H_MF}){]}, yielding
(for an alternative approach to obtain a similar GL functional, see
Ref. \cite{triplet_GL})
\begin{equation}
F=E_{\mathrm{cond}}-\frac{\upsilon}{2}\int_{k}\log\left[\det\left(-\hat{\mathcal{G}}_{k}^{-1}\right)\right],
\end{equation}
where the Green's function is given by $\hat{\mathcal{G}}_{k}^{-1}=\mathrm{i}\omega_{n}-\hat{H}_{\boldsymbol{k}}$,
$\omega_{n}=(2n+1)\pi T$ is a fermionic Matsubara frequency ($n=0,\pm1,\pm2,\cdots$),
the determinant is over the Balian-Werthamer indices, and $\int_{k}=T\sum_{\omega_{n}}\frac{1}{\upsilon}\sum_{\boldsymbol{k}}$.
For simplicity, we introduced the short notation $k=(\boldsymbol{k},\omega_{n})$.
Performing the matrix operations, we find:
\begin{eqnarray}
f(\boldsymbol{M},\Delta_{\mathrm{s}},\Delta_{\mathrm{t}}) & = & -\int_{k}\log\left[\left(\omega_{n}^{2}+E_{+,\boldsymbol{k}}^{2}\right)\left(\omega_{n}^{2}+E_{-,\boldsymbol{k}}^{2}\right)\right]\nonumber \\
 &  & +\frac{2\boldsymbol{M}^{2}}{V_{\mathrm{m}}}+\frac{2|\Delta_{\mathrm{s}}|^{2}}{V_{\mathrm{s}}}+\frac{2|\Delta_{\mathrm{t}}|^{2}}{V_{\mathrm{t}}},\label{action-freq-momen}
\end{eqnarray}
where $f=F/\upsilon$ is the free energy density and $E_{\pm,\boldsymbol{k}}^{2}=\Gamma_{\boldsymbol{k}}\pm\Omega_{\boldsymbol{k}}$
are the squares of the eigen-energies of the reduced Hamiltonian {[}Eq.
(\ref{reduced-H}){]}, with 
\begin{equation}
\Gamma_{\boldsymbol{k}}=|\Delta_{\mathrm{s}}|^{2}+|\Delta_{\mathrm{t}}|^{2}+\boldsymbol{M}^{2}+\left(\xi_{2,\boldsymbol{k}+\boldsymbol{Q}}^{2}+\xi_{1,\boldsymbol{k}}^{2}\right)/2,
\end{equation}
and 
\begin{eqnarray}
\Omega_{\boldsymbol{k}}^{2} & = & \left[2|\Delta_{\mathrm{s}}||\Delta_{\mathrm{t}}|\hat{\boldsymbol{d}}\cos\alpha_{\mathrm{st}}+\boldsymbol{M}\left(\xi_{2,\boldsymbol{k}+\boldsymbol{Q}}+\xi_{1,\boldsymbol{k}}\right)\right]^{2}\nonumber \\
 &  & +|\Delta_{\mathrm{t}}|^{2}\left(\xi_{1,\boldsymbol{k}}-\xi_{2,\boldsymbol{k}+\boldsymbol{Q}}\right){}^{2}+\frac{1}{4}\left(\xi_{1,\boldsymbol{k}}^{2}-\xi_{2,\boldsymbol{k}+\boldsymbol{Q}}^{2}\right)^{2}\nonumber \\
 &  & +4\boldsymbol{M}^{2}|\Delta_{\mathrm{t}}|^{2}\left[1-\left(\hat{\boldsymbol{M}}\cdot\hat{\boldsymbol{d}}\right)^{2}\right].
\end{eqnarray}

These results agree with those of Ref. \cite{Chubukov09,Fernandes10}
for $\Delta_{t}=0$.\textit{\textcolor{red}{{} }}The self-consistent
equations for the order parameters can be calculated from the stationary
points $\partial f\left(\boldsymbol{M},\Delta_{\mathrm{s}},\Delta_{\mathrm{t}}\right)=0$
or, alternatively, through $\langle\psi_{\boldsymbol{k}}\psi_{\boldsymbol{k}}^{\dagger}\rangle=-T\sum_{\omega_{n}}\hat{\mathcal{G}}_{\boldsymbol{k},i\omega_{n}}$.
The matrix inversion of $\hat{\mathcal{G}}_{k}^{-1}$ is straightforward.
For instance, the triplet SC OP is given by
\begin{equation}
\Delta_{\mathrm{t}}=-V_{\mathrm{t}}\int_{k}\frac{\mathcal{Q}_{M}\boldsymbol{M}\cdot\hat{\boldsymbol{d}}+\mathcal{Q}_{d}}{(\omega_{n}^{2}+E_{+,\boldsymbol{k}}^{2})(\omega_{n}^{2}+E_{-,\boldsymbol{k}}^{2})}\mathrm{,}\label{eq_triplet}
\end{equation}
where $\mathcal{Q}_{M}=\Delta_{\mathrm{s}}(\xi_{1,\boldsymbol{k}}+\xi_{2,\boldsymbol{k}+\boldsymbol{Q}})-2\Delta_{\mathrm{t}}\boldsymbol{M}\cdot\hat{\boldsymbol{d}}$
and $\mathcal{Q}_{d}=\Delta_{\mathrm{t}}(\boldsymbol{M}^{2}-|\Delta_{\mathrm{t}}|^{2}-\omega_{n}^{2}-\xi_{1,\boldsymbol{k}}\xi_{2,\boldsymbol{k}+\boldsymbol{Q}})+\Delta_{\mathrm{t}}^{*}\Delta_{\mathrm{s}}^{2}\mathrm{.}$
It is straightforward to show that, in general, t-SC order does not
spontaneously appear (see also Ref. \cite{triplet_Fe_pnictides}).
For instance, in the equation above, setting $M=\Delta_{\mathrm{s}}=0$
and assuming perfect nesting yields the following linearized equation
for $\Delta_{\mathrm{t}}$

\begin{align}
\Delta_{\mathrm{t}} & =\left(V_{\mathrm{t}}\rho_{F}\right)\Delta_{\mathrm{t}}T_{\mathrm{c},\mathrm{t}}\sum_{n}\int d\xi\frac{\omega_{n}^{2}-\xi^{2}}{\left(\omega_{n}^{2}+\xi^{2}\right)^{2}}\nonumber \\
T_{\mathrm{c},\mathrm{t}} & =\frac{W}{2\,\mathrm{arctanh}\left(\frac{1}{V_{\mathrm{t}}\rho_{F}}\right)}\label{eq:T_=00007Bc,t=00007D}
\end{align}
where $\rho_{F}$ is the density of states and $2W$ is the bandwidth.
Clearly, $T_{\mathrm{c},\mathrm{t}}$ only exists if the triplet pairing
interaction $V_{\mathrm{t}}$ is very large, $V_{\mathrm{t}}>\rho_{F}^{-1}$,
implying that the system by itself would never develop t-SC on its
own. However, Eq. (\ref{eq_triplet}) above shows that, as long as
the perfect nesting condition $\xi_{1,\boldsymbol{k}}\neq-\xi_{2,\boldsymbol{k}+\boldsymbol{Q}}$
is not satisfied (a result previously highlighted in Ref. \cite{triplet_Kyung}),
even if we start with $\Delta_{\mathrm{t}}=0$, the triplet components
$\langle c_{\boldsymbol{k},s}f_{-\boldsymbol{k}-\boldsymbol{Q},s^{\prime}}\rangle\propto|\boldsymbol{M}|\mathcal{Q}_{M}\propto|\boldsymbol{M}|\Delta_{\mathrm{s}}$
are self-consistently generated when both $\boldsymbol{M}$ and $\Delta_{\mathrm{s}}$
are non-zero. Thus, when the $SO(3)$ symmetry is spontaneously broken
and the system undergoes a SC phase transition, the SC state is a
combination of singlet and triplet states, even if $V_{\mathrm{t}}=0$.\emph{
}\textit{\emph{Similar results were previously obtained in Ref. }}\emph{\cite{triplet_organics}.}

Near the tetracritical point, both AFM and SC order parameters are
small, and a Ginzburg-Landau (GL) functional approach is justified.
In this case, we expand $f$ {[}Eq. (\ref{action-freq-momen}){]}
for small $|\boldsymbol{M}|$, $|\Delta_{\mathrm{s}}|$ and $|\Delta_{\mathrm{t}}|$
and obtain
\begin{eqnarray}
\Delta f & \approx & \frac{a_{\mathrm{m}}}{2}\boldsymbol{M}^{2}+\frac{a_{\mathrm{s}}}{2}|\Delta_{\mathrm{s}}|^{2}+\frac{a_{\mathrm{t}}}{2}|\Delta_{\mathrm{t}}|^{2}\nonumber \\
 &  & +\lambda\cos\alpha_{\mathrm{st}}|\Delta_{\mathrm{s}}||\Delta_{\mathrm{t}}|\boldsymbol{M}\cdot\hat{\boldsymbol{d}}\nonumber \\
 &  & +\frac{u_{\mathrm{m}}}{4}\boldsymbol{M}^{4}+\frac{u_{\mathrm{s}}}{4}|\Delta_{\mathrm{s}}|^{4}+\frac{u_{\mathrm{t}}}{4}|\Delta_{\mathrm{t}}|^{4}+\frac{\gamma_{\mathrm{ms}}}{2}\boldsymbol{M}^{2}|\Delta_{\mathrm{s}}|^{2}\nonumber \\
 &  & +\frac{\gamma_{\mathrm{mt}}+2\gamma_{12}[1-(\hat{\boldsymbol{M}}\cdot\hat{\boldsymbol{d}}){}^{2}]}{2}\boldsymbol{M}^{2}|\Delta_{\mathrm{t}}|^{2}\nonumber \\
 &  & +\frac{\gamma_{\mathrm{st}}-2\gamma_{12}\sin^{2}\alpha_{\mathrm{st}}}{2}|\Delta_{\mathrm{s}}|^{2}|\Delta_{\mathrm{t}}|^{2}\mathrm{,}\label{action-expanded-final}
\end{eqnarray}
where $\Delta f=f-f(0,0,0)$. The microscopic expressions for the
GL coefficients in terms of the dispersions $\xi_{1,\boldsymbol{k}}$
and $\xi_{2,\boldsymbol{k}+\boldsymbol{Q}}$ and the couplings $V_{\mathrm{m}}$,
$V_{\mathrm{s}}$ and $V_{\mathrm{t}}$ are listed in the Appendix~\ref{sec:GL-coefficients}.
\textit{\emph{Such an expression, without the triplet components,
was previously derived for the two band-model in Ref.}} \cite{Fernandes10}. 

\section{Mean-field phase diagram and collective modes }

\subsection{Mean-field analysis of the Ginzburg-Landau functional}

\begin{figure}
\centering{}\includegraphics[width=7cm]{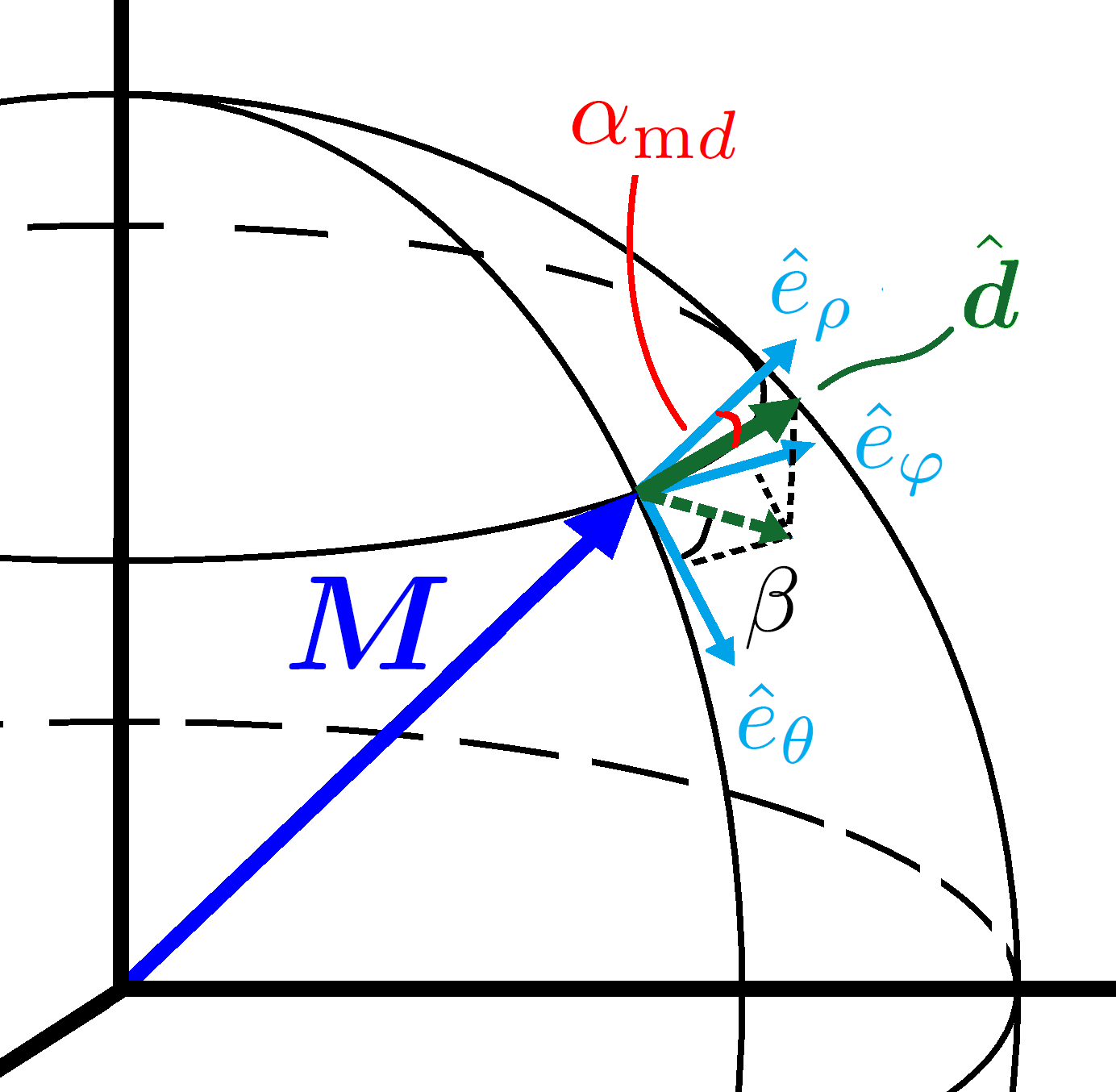}\caption{Schematics of the staggered  magnetization $\boldsymbol{M}$ and the
triplet unit vector $\hat{\boldsymbol{d}}$. We take $\boldsymbol{M}$
to be parallel to $\hat{e}_{\rho}$, $\boldsymbol{M}=M\hat{e}_{\rho}$,
and define the angle between the staggered  magnetization and the
$d$-vector as $\alpha_{\mathrm{m}d}$. The free energy density does
not depend on the angle $\beta$, it only depends on $\hat{\boldsymbol{M}}\cdot\boldsymbol{\hat{d}}=\cos\alpha_{\mathrm{m}d}$.
Therefore, $f$ is invariant with respect to rotations of the $d$-vector
around the staggered  magnetization vector $\boldsymbol{M}$.}
\label{fig:mag-and-d-vec}
\end{figure}

Having derived the Ginzburg-Landau (GL) free energy, Eq. (\ref{action-expanded-final}),
our main goal is to investigate the impact of the t-SC term on the
system's behavior. An obvious consequence of the cubic term coupling
$\Delta_{\mathrm{t}}$, $M$, and $\Delta_{\mathrm{s}}$ is the aforementioned
appearance of t-SC order as soon as antiferromagnetism and singlet
superconductivity coexist, despite the fact that $a_{\mathrm{t}}$
remains positive for all temperatures (i.e. there is no spontaneous
t-SC order). More specifically, minimization of the GL functional
leads to $|\Delta_{\mathrm{t}}|\propto|\boldsymbol{M}||\Delta_{\mathrm{s}}|$.
Thus, because $\Delta_{\mathrm{t}}$ is naturally of second order
in $M$ and $\Delta_{\mathrm{s}}$, we can safely neglect the term
$|\Delta_{\mathrm{t}}|^{4}\propto|\boldsymbol{M}|^{4}|\Delta_{\mathrm{s}}|^{4}$
in the free energy density, as it is effectively of eighth-order. 

We proceed by establishing the phase diagram for the AFM, SC, and
t-SC orders within mean field. First, we express $\boldsymbol{M}$
and $\hat{\boldsymbol{d}}$ in spherical coordinates as $\boldsymbol{M}=M_{\rho}\hat{e}_{\rho}+M_{\theta}\hat{e}_{\theta}+M_{\varphi}\hat{e}_{\varphi}$
and $\hat{\boldsymbol{d}}=d_{\rho}\hat{e}_{\rho}+d_{\theta}\hat{e}_{\theta}+d_{\varphi}\hat{e}_{\varphi}$.
We then set $\boldsymbol{M}=M\hat{e}_{\rho}$ without loss of generality,
as shown in Figure~\ref{fig:mag-and-d-vec}. The angle between $\boldsymbol{M}$
and $\hat{\boldsymbol{d}}$ is denoted $\alpha_{\mathrm{m}d}$, i.e.
$\hat{\boldsymbol{M}}\cdot\boldsymbol{\hat{d}}=d_{\rho}=\cos\alpha_{\mathrm{m}d}$.
Finally, we define $\beta$ as the angle between the projection of
$\hat{\boldsymbol{d}}$\emph{ }\textit{\emph{onto the plane defined
by $\left(\hat{e}_{\theta},\hat{e}_{\varphi}\right)$ and the direction
of}}\textit{ $\hat{e}_{\theta}$,}\textit{\textcolor{red}{{} }}so that
$\hat{d}_{\theta}=\sin\alpha_{\mathrm{m}d}\cos\beta$ and $\hat{d}_{\varphi}=\sin\alpha_{\mathrm{m}d}\sin\beta$
(see Figure~\ref{fig:mag-and-d-vec}).

It is useful to introduce a nine-dimensional ``super-vector'' that
contains all the OPs, corresponding to the amplitude and phase of
each of the $2$ SC order parameters, the $2$ angles characterizing
the unit $d$-vector, and the $3$ components of the magnetization.
In our coordinate system, $\phi^{T}=[M_{\rho},|\Delta_{\mathrm{s}}|,|\Delta_{\mathrm{t}}|,\alpha_{\mathrm{st}},\alpha_{\mathrm{m}d},M_{\theta},M_{\varphi},\alpha_{\mathrm{s}},\beta]$.
We also define the Hessian matrix $\mathbb{H}_{i,j}=\frac{\partial^{2}F}{\partial\phi^{i}\partial\phi^{j}}$
and write the free energy density close to its extremum as
\begin{equation}
\Delta f[\phi^{i}]=\Delta f[\phi_{0}^{i}]+\frac{1}{2}\delta\phi^{T}(\mathbb{H})_{\{\phi_{0}^{i}\}}\delta\phi\mathrm{,}\label{F-close-to-minimum}
\end{equation}
where $\phi_{0}^{T}=[M_{\rho\,0},\cdots,\beta_{0}]$ is the set of
variables at which the first derivatives vanish, i.e., $(\partial_{\phi^{i}}f)_{\{\phi_{0}^{j}\}}=0$.
At the local minimum the Hessian matrix must be positive definite.

The first derivatives of Eq. (\ref{action-expanded-final}) with respect
to the angles $\alpha_{\mathrm{st}}$ and $\alpha_{\mathrm{m}d}$
are given by\begin{widetext}
\begin{equation}
\frac{\partial f}{\partial\alpha_{\mathrm{st}}}=-\sin\alpha_{\mathrm{st}}|\Delta_{\mathrm{s}}||\Delta_{\mathrm{t}}|(\lambda M\cos\alpha_{\mathrm{m}d}+2\gamma_{12}|\Delta_{\mathrm{s}}||\Delta_{\mathrm{t}}|\cos\alpha_{\mathrm{st}})\mathrm{,}\label{df/dast}
\end{equation}
and
\begin{equation}
\frac{\partial f}{\partial\alpha_{\mathrm{m}d}}=-\sin\alpha_{\mathrm{m}d}M|\Delta_{\mathrm{t}}|(\lambda|\Delta_{\mathrm{s}}|\cos\alpha_{\mathrm{st}}-2\gamma_{12}M|\Delta_{\mathrm{t}}|\cos\alpha_{\mathrm{m}d})\mathrm{,}\label{df/damd}
\end{equation}
\end{widetext}respectively. Clearly, a possible solution is $\sin\alpha_{\mathrm{st}}=\sin\alpha_{\mathrm{m}d}=0$,
which is accomplished by $\alpha_{\mathrm{m}d\,0}=\alpha_{\mathrm{st}\,0}=0$
or $\alpha_{\mathrm{m}d\,0}=\alpha_{\mathrm{st}\,0}=\pi$. However,
these solutions do not correspond to a local minimum of the free energy
because, in these cases, since $\lambda>0$, both $\frac{\partial^{2}f}{\partial\alpha_{\mathrm{st}\,0}^{2}}$
and $\frac{\partial^{2}f}{\partial\alpha_{\mathrm{m}d\,0}^{2}}$ are
eigenvalues of the Hessian matrix and negative. The other options
are $\alpha_{\mathrm{m}d\,0}=0$ and $\alpha_{\mathrm{st}\,0}=\pi$
or $\alpha_{\mathrm{m}d\,0}=\pi$ and $\alpha_{\mathrm{st}\,0}=0$.
In these cases, $\partial_{|\Delta_{\mathrm{t}\,0}|}f=0$ gives
\begin{equation}
|\Delta_{\mathrm{t}\,0}|=\frac{\lambda M_{0}|\Delta_{\mathrm{s}\,0}|}{a_{\mathrm{t}}+\gamma_{\mathrm{st}}|\Delta_{\mathrm{s}\,0}|^{2}+\gamma_{\mathrm{mt}}M_{0}^{2}}\mathrm{.}\label{eq:deltat0}
\end{equation}
Imposing now $\partial_{M_{0}}f=0$ and $\partial_{|\Delta_{\mathrm{s}\,0}|}f=0$
and plugging the expression above into the resulting equations leads
to three different solutions with $M_{0}\neq0$ and/or $|\Delta_{\mathrm{s}\,0}|\neq0$:

(i) A pure singlet SC phase with $|\Delta_{\mathrm{s}\,0}|^{2}=-a_{\mathrm{s}}/u_{\mathrm{s}}$,
$M_{0}=0$ and $\Delta_{\mathrm{t}\,0}=0$. The free energy density
for this solution is $f_{\mathrm{s}}=-a_{\mathrm{s}}^{2}/4u_{\mathrm{s}}$;

(ii) A pure AFM phase with $M_{0}^{2}=-a_{\mathrm{m}}/u_{\mathrm{m}}$,
$\Delta_{\mathrm{s}\,0}=0$ and $\Delta_{\mathrm{t}\,0}=0$ whose
condensation energy density is given by $f_{\mathrm{m}}=-a_{\mathrm{m}}^{2}/4u_{\mathrm{m}}$;

(iii) Coexistence of antiferromagnetism and superconductivity where\begin{widetext}
\begin{equation}
a_{\mathrm{m}}+u_{\mathrm{m}}M_{0}^{2}+\left(\gamma_{\mathrm{ms}}-\frac{\lambda^{2}}{a_{\mathrm{t}}+\gamma_{\mathrm{st}}|\Delta_{\mathrm{s}\,0}|^{2}+\gamma_{\mathrm{mt}}M_{0}^{2}}\right)|\Delta_{\mathrm{s}\,0}|^{2}+\frac{\gamma_{\mathrm{mt}}\lambda^{2}M_{0}^{2}|\Delta_{\mathrm{s}\,0}|^{2}}{(a_{\mathrm{t}}+\gamma_{\mathrm{st}}|\Delta_{\mathrm{s}\,0}|^{2}+\gamma_{\mathrm{mt}}M_{0}^{2})^{2}}=0,
\end{equation}
 and 
\begin{equation}
a_{\mathrm{s}}+u_{\mathrm{s}}|\Delta_{\mathrm{s}\,0}|^{2}+\left(\gamma_{\mathrm{ms}}-\frac{\lambda^{2}}{a_{\mathrm{t}}+\gamma_{\mathrm{st}}|\Delta_{\mathrm{s}\,0}|^{2}+\gamma_{\mathrm{mt}}M_{0}^{2}}\right)M_{0}^{2}+\frac{\gamma_{\mathrm{st}}\lambda^{2}M_{0}^{2}|\Delta_{\mathrm{s}\,0}|^{2}}{(a_{\mathrm{t}}+\gamma_{\mathrm{st}}|\Delta_{\mathrm{s}\,0}|^{2}+\gamma_{\mathrm{mt}}M_{0}^{2})^{2}}=0.
\end{equation}
\end{widetext}The solution to these equations and the corresponding
free energy density $f_{\mathrm{coex}}$ can be obtained numerically.

In addition to the solution $\sin\alpha_{\mathrm{st}\,0}=\sin\alpha_{\mathrm{m}d\,0}=0$
for Eqs. (\ref{df/dast}) and (\ref{df/damd}), the conditions $\partial_{\alpha_{\mathrm{st\,0}}}f=0$
and $\partial_{\alpha_{\mathrm{m}d\,0}}f=0$ can also be satisfied
when 
\begin{equation}
2\gamma_{12}|\Delta_{\mathrm{s}\,0}||\Delta_{\mathrm{t}\,0}|\cos\alpha_{\mathrm{st}\,0}=-\lambda M_{0}\cos\alpha_{\mathrm{m}d\,0}\label{alphast0alt}
\end{equation}
and 
\begin{equation}
2\gamma_{12}M_{0}|\Delta_{\mathrm{t}\,0}|\cos\alpha_{\mathrm{m}d\,0}=\lambda|\Delta_{\mathrm{s}\,0}|\cos\alpha_{\mathrm{st}\,0}.\label{alphamd0alt}
\end{equation}
For the two-band model and the microscopic parameters we are considering
(see below), however, we show in Appendix~\ref{sec:Minimization}
that only $\sin\alpha_{\mathrm{st}\,0}=\sin\alpha_{\mathrm{m}d\,0}=0$
is a physical solution corresponding to a minimum of $f$. It follows
that the staggered magnetization $\boldsymbol{M}$ is always parallel
or anti-parallel to the $d$-vector and the relative phase $\alpha_{\mathrm{st}}$
between the singlet and triplet SC order parameters is either zero
or $\pi$.

\begin{figure}
\begin{centering}
\includegraphics[scale=0.5]{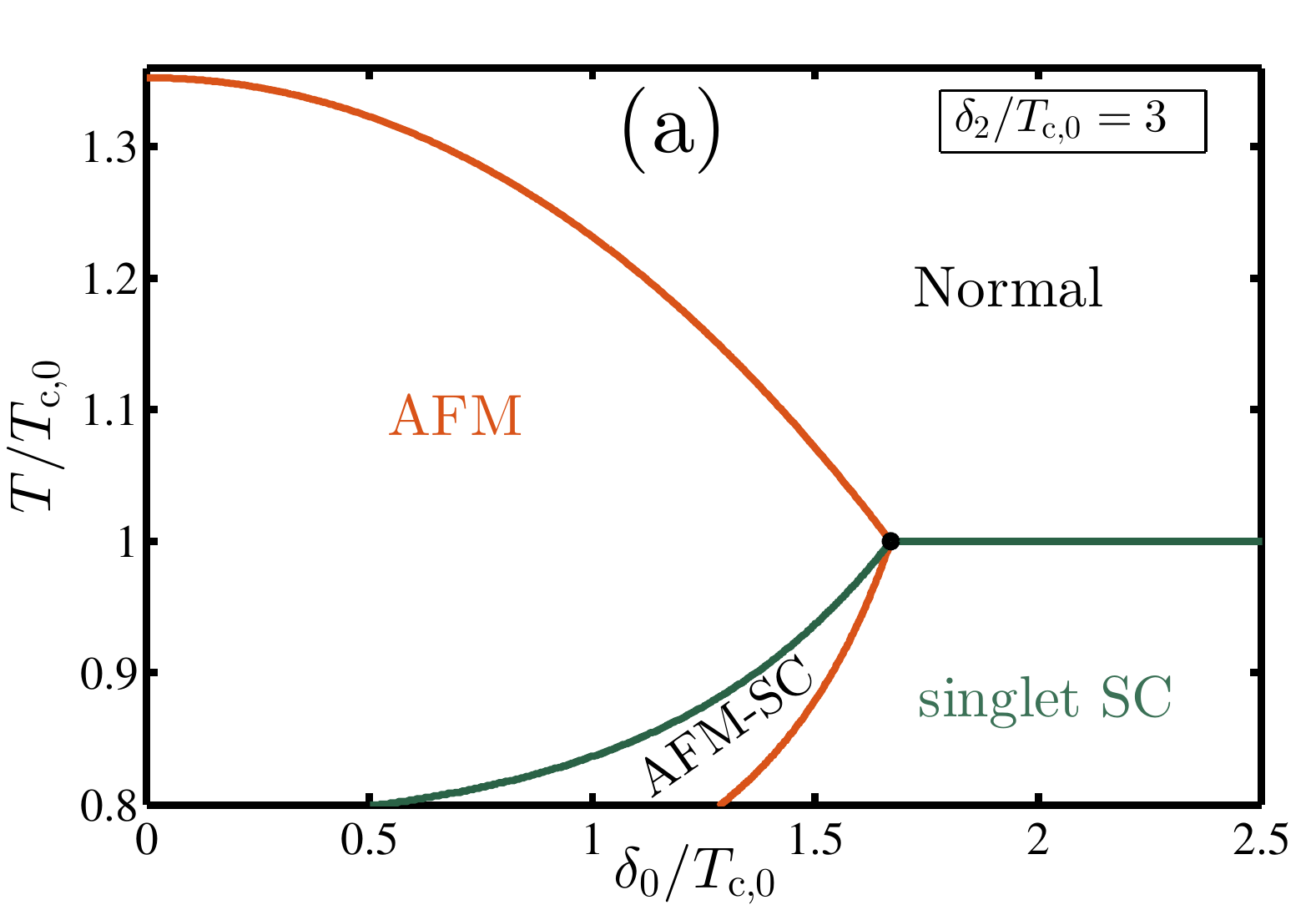}
\par\end{centering}
\centering{}\includegraphics[scale=0.5]{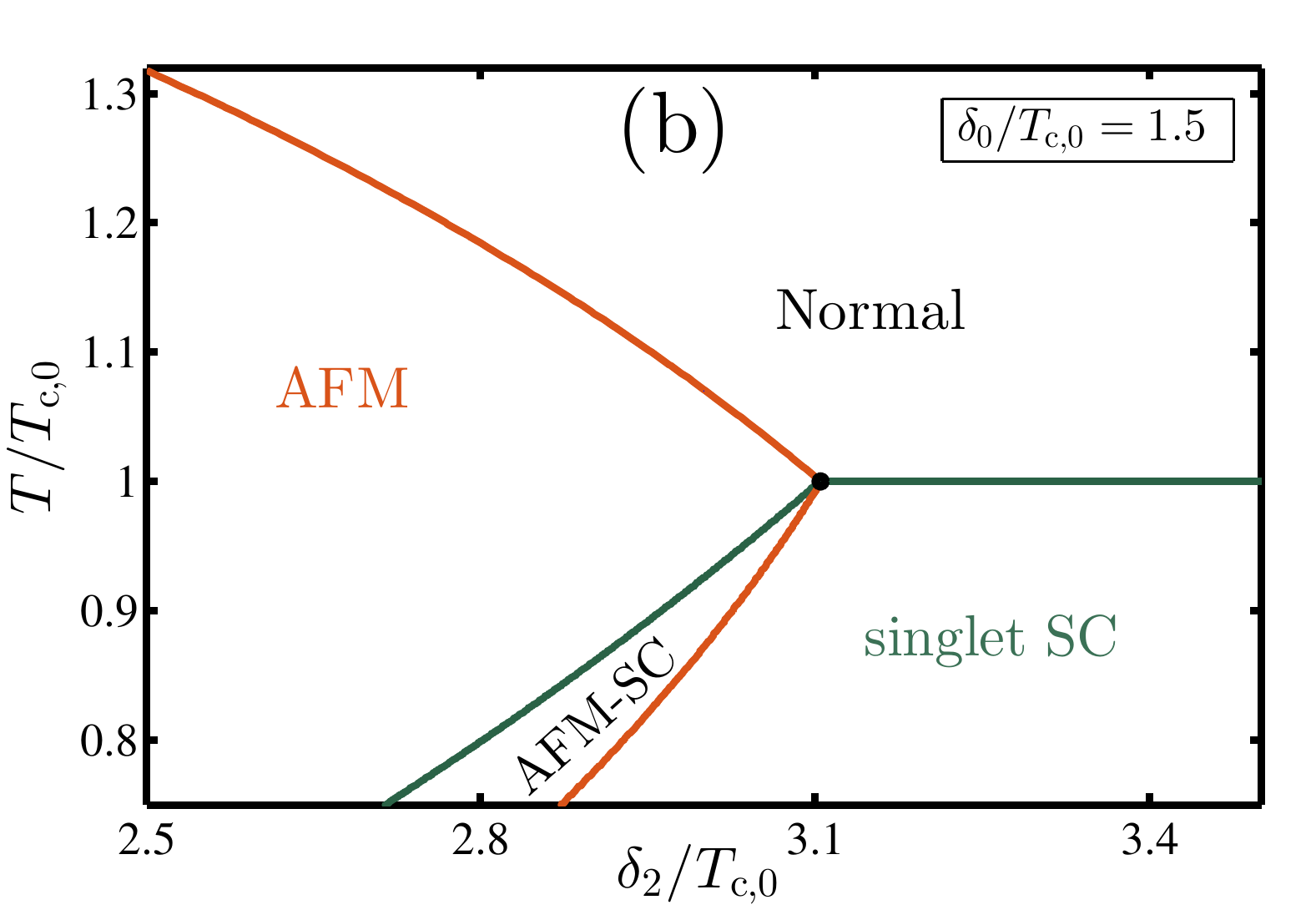}\caption{Phase diagram in the (a) ($T,\delta_{0}$) and (b) ($T,\delta_{2}$)
planes. The green (orange) curve is the singlet SC (AFM) critical
temperature. $M_{0}$ and $\Delta_{\mathrm{s}\,0}$ are both non zero
in the coexistence region located between the green and the orange
curves. Therefore, in this region, $|\Delta_{\mathrm{t}\,0}|\propto M_{0}|\Delta_{\mathrm{s}\,0}|$
is also non-zero. The black dots denote the tetracritical points.
Their coordinates are (a) $(\delta_{0}^{*},T^{*})=(1.669,1)~T_{\mathrm{c},0}$
and (b) $(\delta_{2}^{*},T^{*})=(3.105,1)~T_{\mathrm{c},0}$.}
\label{fig:phase-diagram}
\end{figure}

This is as far as we can go phenomenologically. In our case, however,
the GL parameters are derived directly from the microscopic band dispersions
and interactions, as discussed in Appendix~\ref{sec:GL-coefficients}.
These microscopic parameters are set in the following way: momenta
are measured in units of $k_{F}$ and the Fermi energy $\xi_{\mathrm{F}}\equiv k_{\mathrm{F}}^{2}/2m=\varepsilon_{1,0}-\mu$
is chosen to be $\xi_{\mathrm{F}}=100~\mathrm{meV}$, which gives
$m=0.005~\mathrm{meV}^{-1}$. For the interactions, we used $V_{\mathrm{s}}=266~\mathrm{meV}$,
so that the mean field SC transition temperature \emph{in the absence
of magnetic order} $T_{\mathrm{c},0}=1~\mathrm{meV}$ ($\sim12~\mathrm{K}$),
and $V_{\mathrm{t}}\approx0.1V_{\mathrm{s}}$ (so that $a_{\mathrm{t}}=0.2~\mathrm{meV}^{-1}$).
We also set $V_{\mathrm{m}}=311~\mathrm{meV}$ so that the magnetic
ordering temperature \emph{at perfect nesting and in the absence of
SC} $\bar{T}_{\mathrm{N},0}=2T_{\mathrm{c},0}$. With these parameters
fixed, only two band parameters are left: $\delta_{0}(k)$, which
describes the difference between the areas of hole and electron pockets,
and $\delta_{2}(k)$, which describes the ellipticity of the electron
Fermi pocket. Following previous works \textit{\textcolor{red}{\cite{Chubukov09,Fernandes10}}},
we consider the limit of small Fermi pockets and evaluate these quantities
at $k_{F}$, i.e. $\delta_{0}\equiv\delta_{0}(k_{F})$ and $\delta_{2}\equiv\delta_{2}\left(k_{F}\right)$.
For a fixed value of $\delta_{2}$, we vary $\delta_{0}$ to mimic
the effect of doping and obtain the phase diagram by calculating the
instability lines of each of the three GL solutions discussed above
and comparing their free energies $f_{\mathrm{s}}$ , $f_{\mathrm{m}}$
and $f_{\mathrm{coex}}$. In all cases considered, we noted that the
GL parameters $u_{\mathrm{s}}$,  $\lambda$, $\gamma_{\mathrm{st}}=2\gamma_{12}$
and $\gamma_{\mathrm{ms}}$ are positive, whereas $\gamma_{\mathrm{mt}}$
is negative. The parameter $\lambda$, on the other hand, is such
that $\mathrm{sign}\left(\lambda\right)=\mathrm{sign}\left(\delta_{0}\right)$.
We will consider only the regime $u_{\mathrm{m}}>0$ because this
results in a second order AFM phase transition. If $u_{\mathrm{m}}<0$,
we need to expand the free energy to at least sixth order and, in
this case, if the sixth-order coefficient is positive the transition
will be first order. More details about the GL coefficients can be
found in Appendix~\ref{sec:GL-coefficients}.

\begin{figure}
\begin{centering}
\includegraphics[scale=0.5]{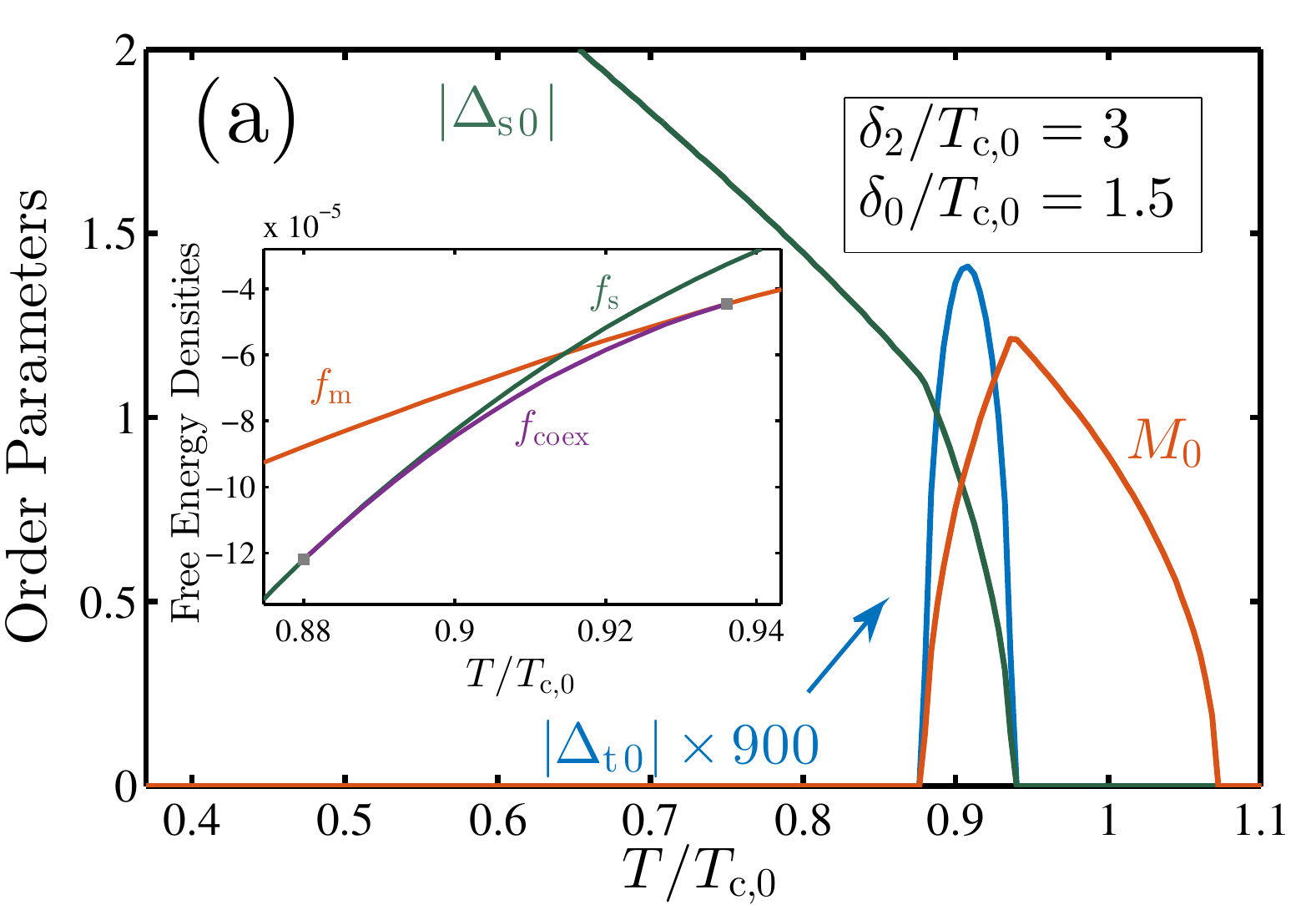}
\par\end{centering}
\begin{centering}
\includegraphics[scale=0.5]{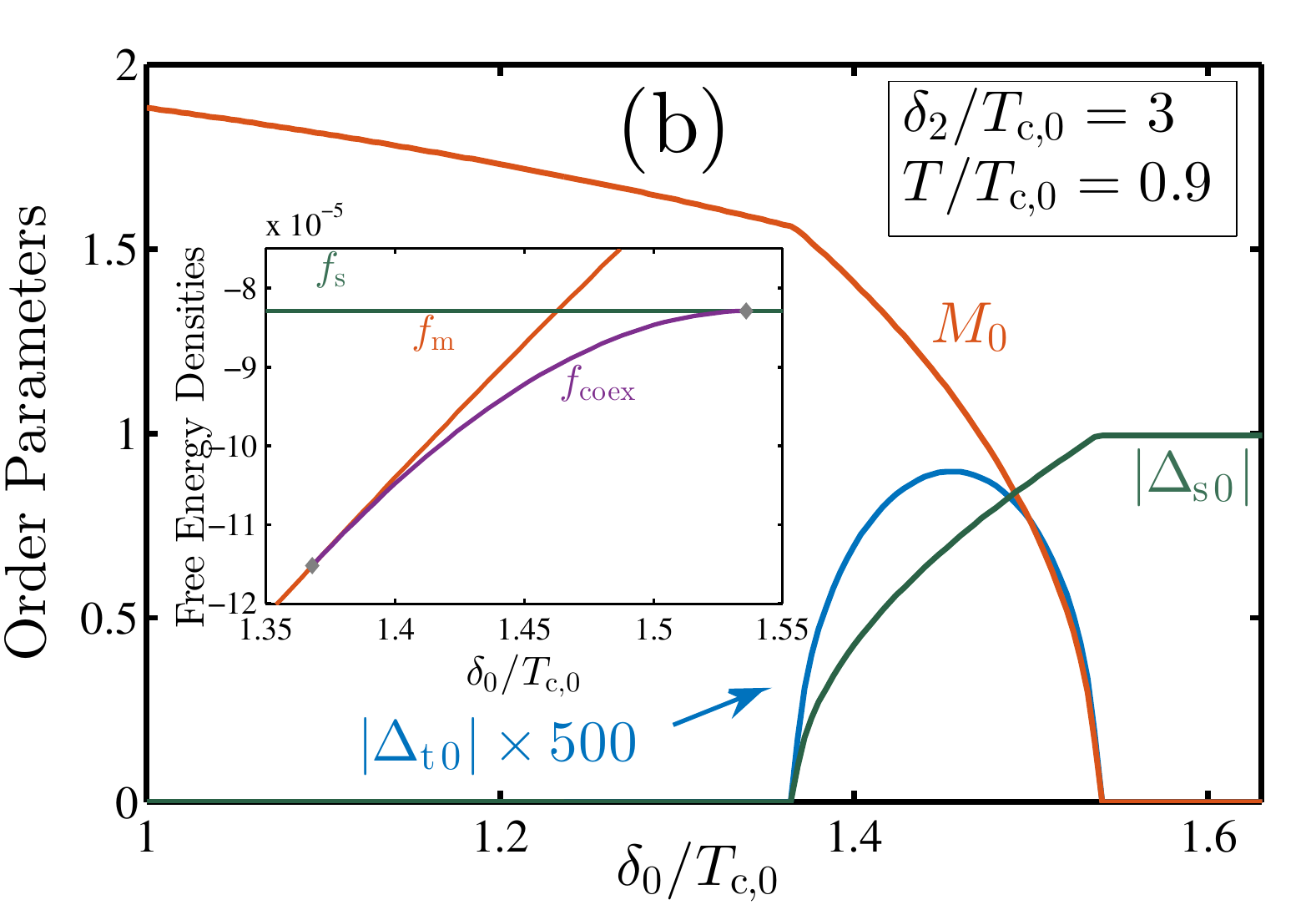}
\par\end{centering}
\centering{}\caption{The behaviors of the singlet SC, triplet SC, and AFM order parameters
($\Delta_{\mathrm{s}\,0}$, $\Delta_{\mathrm{t}\,0}$, and $M_{0}$,
respectively) as function of temperature $T$ {[}(a) fixed $\delta_{0}${]}
and $\delta_{0}$ {[}(b) fixed temperature{]} in the phase diagram
of Fig. \ref{fig:phase-diagram}. The condensation energies of the
pure SC phase $f_{s}$, of the pure magnetic phase $f_{m}$, and of
the AFM-SC coexistence phase $f_{\mathrm{coex}}$ are shown in the
insets. In order to show all quantities in the same plots, we multiplied
$\Delta_{\mathrm{t}\,0}$ by multiplicative factors, as indicated
in the figure.}
\label{fig:order-parameters}
\end{figure}

The phase diagram of the system in the ($T,\delta_{0}$) plane is
shown in Figure~\ref{fig:phase-diagram} for a fixed value of $\delta_{2}$.
Besides the purely AFM and singlet SC phases, there is also the coexistence
phase where both AFM and singlet SC are present, and hence a triplet
SC component as well. This coexistence of SC and AFM is microscopic,
since the $\mathrm{U}(1)$ and $\mathrm{SO}(3)$ symmetries are simultaneously
broken at each and every unit cell of the lattice. In other words,
the lines bounding the AFM-SC region in Figure~\ref{fig:phase-diagram}
are true continuous phase transition lines terminating at a tetracritical
point, and not spinodal lines related to a bicritical point. 

A similar phase diagram, but without the inclusion of triplet SC,
was obtained directly from the microscopic theory in Ref\emph{. }\textit{\emph{\cite{Chubukov09}.
What is the net effect of the t-SC contribution? It turns out that
the }}AFM-SC coexistence phase \emph{expands} when compared to the
case without triplet SC. This stabilizing effect of the triplet pairing
can be understood in simple terms. The smallness of $\Delta_{\mathrm{t}}$
allows us to safely neglect the effectively sixth-order terms $|\Delta_{\mathrm{t}}|^{2}|\Delta_{\mathrm{s}}|^{2}\propto|\boldsymbol{M}|^{2}|\Delta_{\mathrm{s}}|^{4}$
and $|\Delta_{\mathrm{t}}|^{2}|\boldsymbol{M}|^{2}\propto|\boldsymbol{M}|^{4}|\Delta_{\mathrm{s}}|^{2}$
in Eq. (\ref{action-expanded-final}). In this case, Eq.~\eqref{eq:deltat0}
becomes $|\Delta_{\mathrm{t}\,0}|=\lambda M_{0}|\Delta_{\mathrm{s}\,0}|/a_{\mathrm{t}}$.
Eliminating this variable from the free energy density, we obtain
a simplified expression in terms of the AFM and singlet SC OPs only
\begin{eqnarray}
\Delta f_{\mathrm{ms}} & \approx & \frac{a_{\mathrm{m}}}{2}\boldsymbol{M}^{2}+\frac{a_{\mathrm{s}}}{2}|\Delta_{\mathrm{s}}|^{2}+\frac{u_{\mathrm{m}}}{4}\boldsymbol{M}^{4}+\frac{u_{\mathrm{s}}}{4}|\Delta_{\mathrm{s}}|^{4}\nonumber \\
 &  & +\frac{\gamma_{\mathrm{eff}}}{2}\boldsymbol{M}^{2}|\Delta_{\mathrm{s}}|^{2},\label{eq:simpleGL}
\end{eqnarray}
where the effective quartic coupling between $\boldsymbol{M}^{2}$
and $|\Delta_{\mathrm{s}}|^{2}$ is given by 
\begin{equation}
\gamma_{\mathrm{eff}}=\gamma_{\mathrm{ms}}-\lambda^{2}/a_{\mathrm{t}}.\label{eq:gammaeff}
\end{equation}
Thus, we note that the competition between singlet SC and AFM is alleviated
due to the coupling with the t-SC state, as $\gamma_{\mathrm{eff}}<\gamma_{\mathrm{ms}}$,
i.e. the triplet degrees of freedom promote an \emph{effective attraction}
between the AFM and SC order parameters. Evidently, this causes no
changes in the pure singlet SC and AFM solutions.

In Fig. \ref{fig:order-parameters}, we show explicitly the behavior
of the three order parameter, $\Delta_{\mathrm{s}}$, $\Delta_{\mathrm{t}}$,
and $M$, as functions of temperature (for fixed $\delta_{0}/T_{\mathrm{c},0}=1.5$)
and as functions of $\delta_{0}$ (for fixed temperature $T/T_{\mathrm{c},0}=0.9$).
The competition between $\Delta_{\mathrm{s}}$ and $M$ is evident,
as well as the secondary character of the triplet order parameter,
which is much smaller than $\Delta_{\mathrm{s}}$ and $M$. The condensation
energies of each phase are also shown in the insets, highlighting
that the AMF-SC coexistence region is indeed the global energy minimum.

\subsection{Excitations in the AFM-SC coexistence state}

Having shown that the phase diagram contains the AFM-SC coexistence
phase, we now discuss its collective modes by studying the Hessian
matrix $\mathbb{H}_{i,j}$ defined in Eq. \ref{F-close-to-minimum}.
Inspection of Eq.~\eqref{action-expanded-final} reveals that the
free energy is independent of the last 4 components of the super-vector
$\phi$. Therefore, the corresponding $4\times4$ block of $\mathbb{H}_{i,j}$
vanishes identically. Evidently, this reflects (a) the rotational
$\mathrm{SO}(3)$ symmetry of the antiferromagnetic order parameter
$\left(M_{\theta},M_{\varphi}\right)$, (b) the global $\mathrm{U}(1)$
symmetry of the SC order parameter $\left(\alpha_{s}\right)$ and
(c) the fact that the vector $\hat{\boldsymbol{d}}$ can be freely
rotated around the antiferromagnetic order parameter without any energy
cost $\left(\beta\right)$. These symmetries are spontaneously broken
in the ordered phases. There is one Goldstone mode associated with
each one of these variables once the corresponding symmetries are
broken, except for the global SC phase $\alpha_{s}$ which is gapped
out by the coupling to the electromagnetic field through the Anderson-Higgs
mechanism. We will drop this $4\times4$ block in what follows, and
focus on the non-vanishing part of the Hessian matrix in the coexistence
state, given by:

\begin{equation}
\mathbb{H}=\left[\begin{array}{c|cc}
\mathbb{C}_{3\times3} & 0 & 0\\
\hline 0 & \frac{\partial^{2}f}{\partial\alpha_{\mathrm{st}}^{2}} & 0\\
0 & 0 & \frac{\partial^{2}f}{\partial\alpha_{\mathrm{m}d}^{2}}
\end{array}\right],
\end{equation}
where \begin{widetext} 
\begin{equation}
\frac{\partial^{2}f}{\partial\alpha_{\mathrm{st}}^{2}}=\frac{\lambda^{2}M^{2}\Delta_{\mathrm{s}}^{2}}{a_{\mathrm{t}}+\gamma_{\mathrm{st}}\Delta_{\mathrm{s}}^{2}+\gamma_{\mathrm{mt}}M^{2}}\left(1-\frac{2\gamma_{12}\Delta_{\mathrm{s}}^{2}}{a_{\mathrm{t}}+\gamma_{\mathrm{st}}\Delta_{\mathrm{s}}^{2}+\gamma_{\mathrm{mt}}M^{2}}\right),\label{leggmode}
\end{equation}
 and 
\begin{equation}
\frac{\partial^{2}f}{\partial\alpha_{\mathrm{m}d}^{2}}=\frac{\lambda^{2}M^{2}\Delta_{\mathrm{s}}^{2}}{a_{\mathrm{t}}+\gamma_{\mathrm{st}}\Delta_{\mathrm{s}}^{2}+\gamma_{\mathrm{mt}}M^{2}}\left(1+\frac{2\gamma_{12}M^{2}}{a_{\mathrm{t}}+\gamma_{\mathrm{st}}\Delta_{\mathrm{s}}^{2}+\gamma_{\mathrm{mt}}M^{2}}\right).\label{mdmode}
\end{equation}
\end{widetext} 

While the $3\times3$ matrix $\mathbb{C}_{3\times3}$ refers to collective
amplitude modes related to the equilibrium values of $\Delta_{\mathrm{s}}$,
$\Delta_{\mathrm{t}}$, and $M$, the last two quantities refer to
the relative phase between the two SC order parameters, $\alpha_{\mathrm{st}}$,
and to the relative angle between the $d$-vector and the magnetization,
$\alpha_{\mathrm{m}d}$. Although $\mathbb{C}_{3\times3}$ is straightforward
to obtain, we refrain from writing out explicitly its lengthy expression
here. We have scanned exhaustively the values of $\delta_{0}$, $\delta_{2}$,
and $T$ in the AFM-SC coexistence region and found consistently that
the eigenvalues of $\mathbb{C}_{3\times3}$ are indeed always positive,
which proves that we have a locally stable phase. Moreover, as emphasized
before, it is also the global minimum.

As for the terms $\frac{\partial^{2}f}{\partial\alpha_{\mathrm{st}}^{2}}$
and $\frac{\partial^{2}f}{\partial\alpha_{\mathrm{m}d}^{2}}$, we
also found them to be always positive. Specifically, Eq.~\eqref{leggmode}
gives the ``mass'' (i.e. the energy at $\boldsymbol{k}=0$) of the
collective mode associated with oscillations of the relative phase
between the two SC order parameters. It is thus the analogue of the
Leggett mode of two-band SCs \cite{Leggett66}. Similarly, the other
second derivative in Eq.~\eqref{mdmode} gives the ``mass'' of another
collective mode corresponding to oscillations of the angle between
the AFM OP and the $\hat{\boldsymbol{d}}$ vector of the t-PDW. It
is useful to consider the simplified GL functional in Eq.~\eqref{eq:simpleGL},
which was obtained after neglecting the effectively sixth-order terms
coming from the t-SC OP. In this approximation we find 
\begin{equation}
\frac{\partial^{2}f}{\partial\alpha_{\mathrm{st}\,0}^{2}}=\frac{\partial^{2}f}{\partial\alpha_{\mathrm{m}d\,0}^{2}}=\frac{\lambda^{2}}{a_{\mathrm{t}}}M_{0}^{2}\Delta_{\mathrm{s}\,0}^{2}.
\end{equation}
Thus, the masses of the Leggett mode and of the angular mode between
$\boldsymbol{M}$ and $\hat{\boldsymbol{d}}$ become degenerate. As
we have seen above, this degeneracy is lifted with the inclusion of
the sixth-order terms.

\section{Impact of fluctuations on the phase diagram \label{sec:Magnetic-fluctuations}}

In this section we go beyond the previous mean-field analysis and
investigate the impact of Gaussian fluctuations on the phase diagram
of Fig.~\ref{fig:phase-diagram}. Because the SC transition is usually
well described by a mean field transition, we here focus on the impact
of magnetic fluctuations only. In particular, our goal is to determine
how the mean-field critical temperatures ($T_{\mathrm{c},0}$ and
$T_{\mathrm{N},0}$) as well as the coexistence region are affected
by these magnetic fluctuations.

We first generalize the uniform staggered magnetization to an inhomogeneous
function of space $\boldsymbol{M}\to\boldsymbol{M}_{\boldsymbol{x}}$,
or in the Fourier space $\boldsymbol{M}_{\boldsymbol{q}}=\sum_{\boldsymbol{x}}e^{\mathrm{i}\boldsymbol{q}\cdot\boldsymbol{x}}\boldsymbol{M}_{\boldsymbol{x}}$.
We assume this extension does not change in a relevant way any coupling
other than the quadratic magnetic coefficient of the free energy {[}Eq.
(\ref{action-expanded-final}){]}, whereby $\frac{a_{\mathrm{m}}}{2}\boldsymbol{M}^{2}\to\frac{a_{\mathrm{m}}+g_{\boldsymbol{q}}}{2}|\boldsymbol{M}_{\boldsymbol{q}}|^{2}$,
where $(a_{\mathrm{m}}+g_{\boldsymbol{q}})^{-1}$ is the momentum-dependent
magnetic susceptibility with $g_{\boldsymbol{q}}$ being some function
of momentum such that $g_{0}=0$. 

We decouple the quartic AFM term in the partition function, $Z\propto\int\mathcal{D}[\boldsymbol{M},\Delta_{\mathrm{s}},\Delta_{\mathrm{t}}]e^{-F/T}$,
via a Hubbard-Stratonovitch transformation \cite{Fernandes12}
\begin{equation}
e^{-\frac{u_{\mathrm{m}}}{4T}\sum_{\boldsymbol{x}}\boldsymbol{M}_{\boldsymbol{x}}^{4}}\propto\int\mathcal{D}[\psi]e^{-\frac{1}{2T}\sum_{\boldsymbol{x}}\left(-\frac{\psi_{\boldsymbol{x}}^{2}}{2u_{\mathrm{m}}}+\boldsymbol{M}_{\boldsymbol{x}}^{2}\psi_{\boldsymbol{x}}\right)}\mathrm{.}
\end{equation}
The price we pay when we introduce the auxiliary Hubbard-Stratonovitch
field is an additional degree of freedom in the partition function
($\mathcal{D}[\psi]$). The effective free energy density thus becomes
quadratic in the magnetic order parameter
\begin{eqnarray}
f_{\mathrm{eff}} & = & \frac{a_{\mathrm{s}}}{2}|\Delta_{\mathrm{s}}|^{2}+\frac{a_{\mathrm{t}}}{2}|\Delta_{\mathrm{t}}|^{2}+\frac{u_{\mathrm{s}}}{4}|\Delta_{\mathrm{s}}|^{4}-\frac{\psi^{2}}{4u_{\mathrm{m}}}\nonumber \\
 &  & +\frac{1}{2\upsilon^{2}}\sum_{\boldsymbol{q}}(g_{\boldsymbol{q}}+a_{\mathrm{m}}+\gamma_{\mathrm{ms}}|\Delta_{\mathrm{s}}|^{2}+\psi)|\boldsymbol{M}_{\boldsymbol{q}}|^{2}\nonumber \\
 &  & +\frac{1}{\upsilon^{2}}\sum_{\boldsymbol{q}}\lambda\cos\alpha_{\mathrm{st}}|\Delta_{\mathrm{s}}||\Delta_{\mathrm{t}}|\boldsymbol{M}_{\boldsymbol{q}}\cdot\boldsymbol{d}_{-\boldsymbol{q}}\mathrm{,}\label{F_eff-mag-fluctuations}
\end{eqnarray}
where $\hat{\boldsymbol{d}}_{\boldsymbol{q}}=\sum_{\boldsymbol{x}}e^{\mathrm{i}\boldsymbol{q}\cdot\boldsymbol{x}}\hat{\boldsymbol{d}}=\upsilon\delta_{\boldsymbol{q},0}\hat{\boldsymbol{d}}$
and we have neglected the sixth-order terms $|\Delta_{\mathrm{t}}|^{4}$,
$|\Delta_{\mathrm{t}}|^{2}|\Delta_{\mathrm{s}}|^{2}$ and $|\Delta_{\mathrm{t}}|^{2}|\boldsymbol{M}_{\boldsymbol{q}}|^{2}$.
Note that we assumed $\psi_{\boldsymbol{x}}$ to be homogeneous, which
can be justified in the saddle-point approximation that corresponds
to evaluating the partition function at $\partial f_{\mathrm{eff}}(\psi)/\partial\psi=0$.
At this saddle point, $\psi=u_{m}\left\langle M^{2}\right\rangle $
is proportional to the Gaussian magnetic fluctuations. The saddle
point can be justified in an appropriate large-$N$ limit of a theory
in which the number of components of $\boldsymbol{M}$ is enlarged
from $3\to N$. The integration over the $\Delta_{\mathrm{t}}$ fields
can always be done, in any state, because according to Eq.~\eqref{eq:T_=00007Bc,t=00007D}
the field is always massive, i.e. $a_{\mathrm{t}}>0$ at all temperatures.
Then, we introduce $\Delta_{\mathrm{t}}\boldsymbol{d}=\boldsymbol{\Delta}_{\mathrm{t}}$
and integrate over $\Delta_{\mathrm{t},j}$ to obtain
\begin{eqnarray}
f_{\mathrm{eff}} & = & \frac{a_{\mathrm{s}}}{2}|\Delta_{\mathrm{s}}|^{2}+\frac{u_{\mathrm{s}}}{4}|\Delta_{\mathrm{s}}|^{4}-\frac{\psi^{2}}{4u_{\mathrm{m}}}\nonumber \\
 &  & +\frac{1}{2\upsilon^{2}}\sum_{\boldsymbol{q}}(g_{\boldsymbol{q}}+a_{\mathrm{m}}+\gamma_{\mathrm{eff}}|\Delta_{\mathrm{s}}|^{2}+\psi)|\boldsymbol{M}_{\boldsymbol{q}}|^{2}\mathrm{,}\label{F_eff-apos-integrar-trip}
\end{eqnarray}
where $\gamma_{\mathrm{eff}}$ was defined in Eq.~\eqref{eq:gammaeff}. 

We also introduce magnetic long-range order by allowing the radial
component of $\boldsymbol{M}_{\boldsymbol{q}}$ to have a nonzero
mean value. We write $M_{\boldsymbol{q}}^{(\rho)}\to\upsilon M\delta_{\boldsymbol{q},0}+(1-\delta_{\boldsymbol{q},0})M_{\boldsymbol{q}}^{(\rho)}$
and integrate out the magnetic fluctuations $M_{\boldsymbol{q}}^{(j)}$
to obtain
\begin{eqnarray}
f_{\mathrm{eff}} & = & \frac{a_{\mathrm{s}}}{2}|\Delta_{\mathrm{s}}|^{2}+\frac{u_{\mathrm{s}}}{4}|\Delta_{\mathrm{s}}|^{4}-\frac{\psi^{2}}{4u_{\mathrm{m}}}+\frac{r}{2}M^{2}\nonumber \\
 &  & +\frac{NT}{2\upsilon}\sum_{\boldsymbol{q}}\log(g_{\boldsymbol{q}}+r)-\frac{T}{2\upsilon}\log\,r\mathrm{,}\label{F_eff-apos-integrar-flut-mag}
\end{eqnarray}
where $r=a_{\mathrm{m}}+\gamma_{\mathrm{eff}}|\Delta_{\mathrm{s}}|^{2}+\psi$
is the ``mass'' of the fully renormalized susceptibility. In order
to extended the number of components of the staggered magnetization
from $N=3$ to arbitrary $N$ we have to rescale the OP and the couplings
as $(M^{2},\Delta_{\mathrm{s}}^{2})\to(M^{2},\Delta_{\mathrm{s}}^{2})N$
and $(u_{\mathrm{s}},u_{\mathrm{m}},\gamma_{\mathrm{eff}})\to(u_{\mathrm{s}},u_{\mathrm{m}},\gamma_{\mathrm{eff}})/N$.
The effectivefree energy density then reads
\begin{eqnarray}
f_{\mathrm{eff}}/N & = & \frac{a_{\mathrm{s}}}{2}|\Delta_{\mathrm{s}}|^{2}+\frac{u_{\mathrm{s}}}{4}|\Delta_{\mathrm{s}}|^{4}-\frac{\psi^{2}}{4u_{\mathrm{m}}}+\frac{r}{2}M^{2}\nonumber \\
 &  & +\frac{T_{\mathrm{c},0}}{2\upsilon}\sum_{\boldsymbol{q}}\log(g_{\boldsymbol{q}}+r)\mathrm{,}\label{F_eff-final}
\end{eqnarray}
for $N\gg1$. In the spirit of the GL approximation we have set $T\approx T_{\mathrm{c},0}$
in the last term of the above equation. 

Extremizing $f_{\mathrm{eff}}$ with respect to the Hubbard-Stratonovitch
field $\psi$ leads to the following equation

\begin{equation}
\frac{\psi}{u_{\mathrm{m}}}=M^{2}+\mathcal{I}(r)\mathrm{,}\label{min_r}
\end{equation}
where
\begin{equation}
\mathcal{I}(r)=\frac{T_{\mathrm{c},0}}{\upsilon}\sum_{\boldsymbol{q}}\frac{1}{g_{\boldsymbol{q}}+r}\mathrm{.}\label{def-I(r)}
\end{equation}
On the other hand, extremizing $f_{\mathrm{eff}}$ with respect to
the order parameters $M$ and $|\Delta_{\mathrm{s}}|$ we obtain
\begin{equation}
M=0\quad\mathrm{or}\quad r=0\label{min_M}
\end{equation}
and
\begin{equation}
|\Delta_{\mathrm{s}}|=0\quad\mathrm{or}\quad a_{\mathrm{s}}+u_{\mathrm{s}}|\Delta_{\mathrm{s}}|^{2}+\frac{\gamma_{\mathrm{eff}}\psi}{u_{\mathrm{m}}}=0\mathrm{,}\label{min_Delta_s}
\end{equation}
respectively. The set of Eqs.~(\ref{min_r}) - (\ref{min_Delta_s})
has four different solutions, as in the case without magnetic fluctuations.
The possible phases are:

(i) A pure singlet SC phase with $M=0$, $\psi=u_{\mathrm{m}}\mathcal{I}(a_{\mathrm{m}}+\gamma_{\mathrm{eff}}|\Delta_{\mathrm{s}}|^{2}+\psi)$
and
\begin{equation}
|\Delta_{\mathrm{s}}|^{2}=-\frac{a_{\mathrm{s}}}{u_{\mathrm{s}}}-\frac{\gamma_{\mathrm{eff}}\psi}{u_{\mathrm{s}}u_{\mathrm{m}}};\label{Delta_s-SC-phase}
\end{equation}

(ii) A pure AFM phase with $\Delta_{\mathrm{s}}=0$, $\psi=-a_{\mathrm{m}}$
and
\begin{equation}
M^{2}=-\frac{a_{\mathrm{m}}}{u_{\mathrm{m}}}-\mathcal{I}(0)\mathrm{;}\label{M-AFM-phase}
\end{equation}

(iii) A phase of coexistence of AFM and SC with $r=0$,
\begin{equation}
|\Delta_{\mathrm{s}}|^{2}=\frac{a_{\mathrm{m}}\gamma_{\mathrm{eff}}-a_{\mathrm{s}}u_{\mathrm{m}}}{\gamma_{\mathrm{eff}}^{2}-u_{\mathrm{s}}u_{\mathrm{m}}}\label{Delta_s-AFM+SC-phase}
\end{equation}
and
\begin{equation}
M^{2}=\frac{a_{\mathrm{s}}\gamma_{\mathrm{eff}}-a_{\mathrm{m}}u_{\mathrm{s}}}{\gamma_{\mathrm{eff}}^{2}-u_{\mathrm{s}}u_{\mathrm{m}}}-\mathcal{I}(0)\mathrm{;}\label{r-AFM+SC-phase}
\end{equation}

(iv) The normal state with $\Delta_{\mathrm{s}}=M=0$ and $r-a_{\mathrm{m}}=u_{\mathrm{m}}\mathcal{I}(r)$.

The quantity $\mathcal{I}(0)$ in Eq.~\eqref{M-AFM-phase} measures
the change in the AFM critical temperature $T_{\mathrm{N},0}$ due
to the Gaussian AFM fluctuations. Since $\mathcal{I}(r)\geqslant0$,
we conclude that $T_{\mathrm{N},0}$ is suppressed by magnetic fluctuations,
as expected. We can also see from Eq.~\eqref{def-I(r)} that, in
a two-dimensional system, the magnetic fluctuations correction diverges
($\mathcal{I}(0)\to\infty$), thus destroying the magnetic order.
This is a consequence of the Mermin-Wagner theorem \cite{Mermin-Wagner},
which states that a finite-temperature AFM transition only happens
for dimensions $d>2$. We will, therefore, consider an anisotropic
three-dimensional model of weakly coupled layers, for which \cite{Fernandes12}
\begin{equation}
g_{\boldsymbol{q}}=\kappa(q_{x}^{2}+q_{y}^{2})+\eta_{z}\sin^{2}(q_{z}/2)\label{mag_susceptibility}
\end{equation}
with $0\leqslant q_{z}<2\pi$ and $\eta_{z}<\kappa$. A detailed derivation
of the microscopic expression for $\kappa$ can be found in Appendix~\ref{sec:GL-coefficients}.
Carrying out the calculations we obtain
\begin{equation}
\mathcal{I}(r)=\frac{T_{\mathrm{c},0}}{2\pi\kappa}\,\log\left(\frac{\sqrt{r+\kappa\Lambda^{2}}+\sqrt{r+\kappa\Lambda^{2}+\eta_{z}}}{\sqrt{r}+\sqrt{r+\eta_{z}}}\right)\mathrm{,}\label{I(r)--final}
\end{equation}
where $\Lambda$ is an ultra-violet cutoff \cite{Fernandes12}. For
completeness, we also show the result of the momentum summation in
the last term of the effective free energy (\ref{F_eff-final})
\begin{eqnarray}
\mathcal{I}_{\ell}(r) & \equiv & \frac{1}{\upsilon}\sum_{\boldsymbol{q}}\log(g_{\boldsymbol{q}}+r)\nonumber \\
 & = & \frac{\Lambda^{2}}{2\pi}\log\left(\sqrt{r+\kappa\Lambda^{2}}+\sqrt{r+\kappa\Lambda^{2}+\eta_{z}}\right)\nonumber \\
 &  & +\frac{\eta_{z}+2r}{4\pi\kappa}\log\left(\frac{\sqrt{r+\kappa\Lambda^{2}}+\sqrt{r+\kappa\Lambda^{2}+\eta_{z}}}{\sqrt{r}+\sqrt{r+\eta_{z}}}\right)\nonumber \\
 &  & +\frac{\sqrt{r+\kappa\Lambda^{2}}-\sqrt{r+\kappa\Lambda^{2}+\eta_{z}}}{4\pi\kappa}+\mathrm{constants.}\label{I_log}
\end{eqnarray}

\begin{figure}
\begin{centering}
\includegraphics[width=1\columnwidth]{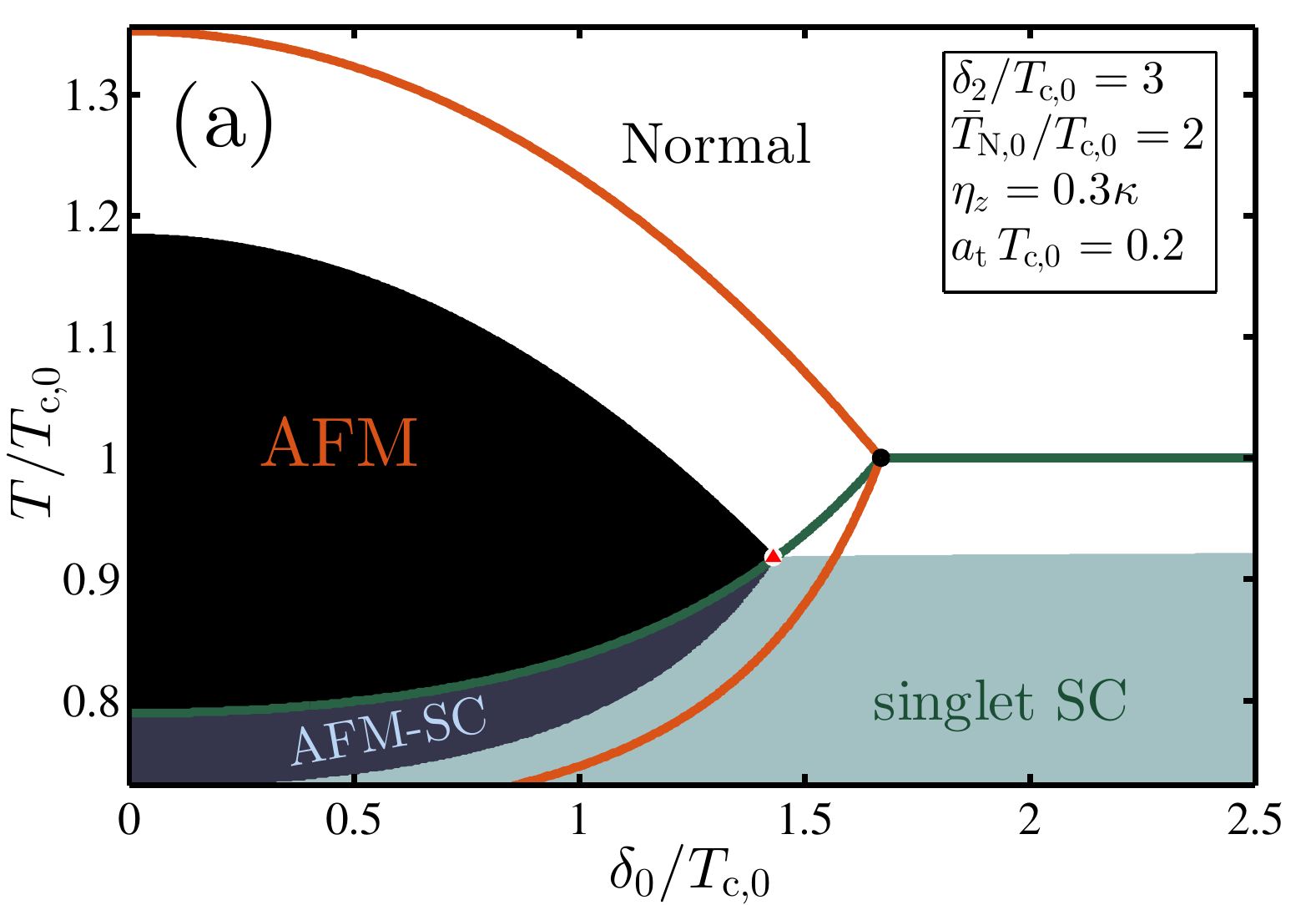}
\par\end{centering}
\begin{centering}
\includegraphics[width=1\columnwidth]{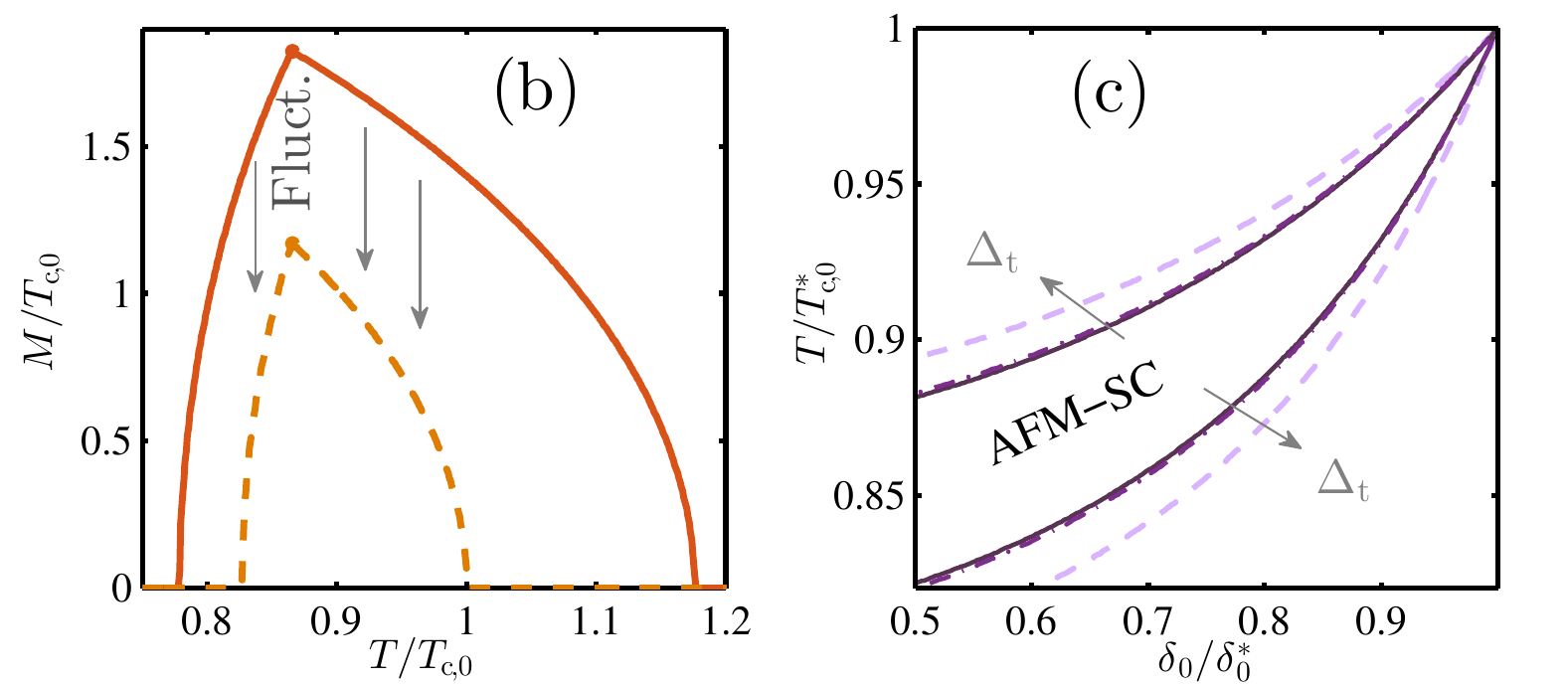}
\par\end{centering}
\centering{}\caption{(a) Fluctuation-corrected phase diagram in the ($T,\delta_{0}$) plane.
The black, light gray, dark gray and white regions are the pure AFM,
pure SC, coexistence AFM-SC, and normal phases, respectively. The
green and orange curves represent the phase diagram without fluctuations
{[}same lines as in Fig.~\ref{fig:phase-diagram}(a){]}. Clearly,
magnetic fluctuations shrink the AFM region. The ``new'' multicritical
point, represented by the red triangle, is still a tetracritical point.
(b) The AFM order parameter for the same parameters of panel (a) and
$\delta_{0}=1.20\,T_{\mathrm{c},0}$ with (dashed line) and without
(solid line) the inclusion of magnetic fluctuations. The solid circles
denote the positions of the SC critical temperatures. (c) The effect
of the triplet SC order parameter on the boundaries of the coexistence
region: the greater the value of $a_{\mathrm{t}}$, the smaller $\Delta_{\mathrm{t}}$
becomes. Solid curves are for $a_{\mathrm{t}}\to\infty$ ($\Delta_{\mathrm{t}}\equiv0$),
dash-dotted curves for $a_{\mathrm{t}}=0.2\,\mathrm{meV}^{-1}$ {[}as
in panel (a){]}, and dashed curves for $a_{\mathrm{t}}=0.02\,\mathrm{meV}^{-1}$.\label{phase-diagram-fluctuations}}
\end{figure}

We solved the set of coupled nonlinear equations for $r$, $|\Delta_{\mathrm{s}}|^{2}$
and $M^{2}$ and compared the values of the free energies of the possible
phases to obtain the fluctuation-corrected phase diagram of the model,
as shown in Fig.~\ref{phase-diagram-fluctuations}(a). We set $\eta_{z}=0.3\kappa$
and $\Lambda^{2}=k_{\mathrm{F}}^{2}=10$. The shaded areas represent
the fluctuation-corrected phases, whereas the full lines represent
the phase transition boundaries in the absence of fluctuations {[}i.e.
the same lines depicted in Fig.~\ref{fig:phase-diagram}(a){]}. We
clearly see that both the mean-field SC and the mean-field Néel critical
temperatures are reduced by the magnetic fluctuations. These suppressions
occur because the last terms of Eqs. \eqref{Delta_s-SC-phase} and
\eqref{M-AFM-phase} are negative, i.e., the OPs are reduced. Analogously,
the effect of the magnetic fluctuations on $M^{2}$ in the coexistence
phase is given by the last term of Eq.~\eqref{r-AFM+SC-phase}, which
is negative. We illustrate the effect of the magnetic fluctuations
on the staggered magnetization in Fig.~\ref{phase-diagram-fluctuations}(b),
which shows that the reduction of $M$ within the AFM-SC coexistence
region implies that the \emph{lower} temperature at which magnetic
order disappears is \emph{enhanced} by the magnetic fluctuations.
Finally, the SC transition temperature in the magnetically ordered
state occurs at same temperature when compared to the case without
magnetic fluctuations. This is evident from Eq. \eqref{Delta_s-AFM+SC-phase},
since the singlet SC OP is not affected by the magnetic fluctuations.

To elucidate the role of the triplet degrees of freedom on the coexistence
phase, we also changed the value of $a_{\mathrm{t}}$, since $|\Delta_{\mathrm{t}}|\propto a_{\mathrm{t}}^{-1}$.
We show the transition lines to the AFM-SC coexistence phase for three
different values of $a_{\mathrm{t}}$ in Fig.~\ref{phase-diagram-fluctuations}(c).\textcolor{red}{\emph{
}}Clearly, the larger the value of $|\Delta_{\mathrm{t}}|$ the larger
the size of the AFM-SC region, thus showing that the stabilizing effect
of the triplet component on the coexistence region is not restricted
to the mean-field analysis of Section III.A, but is also present when
fluctuations are included. The most prominent result of these renormalizations,
therefore, is the evident shrinking of the AFM region caused by the
magnetic fluctuations, which is to be expected. We checked, by comparing
the various free energies, that the phases indicated in Fig.~\ref{phase-diagram-fluctuations}
are indeed the thermodynamically stable phases of the system. Furthermore,
all the phase transition lines keep their second-order character and
their intersection remains a tetracritical point. 

\section{Conclusions}

In this paper, we studied the impact of the spin triplet pairing component
on the phase diagram of competing AFM and SC orders. Except in very
special cases, such as systems with perfectly nested bands, the t-SC
is always present in the AFM-SC coexistence phase, and is therefore
an integral part of the phase diagram of systems displaying these
two types of order. As we showed, in general the triplet degrees of
freedom suppress the competition between AFM and SC by mediating an
effective attraction between these otherwise competing orders. More
importantly, we investigated in detail the coupling between the triplet
$d$-vector and the staggered magnetization. In the ordered state,
this coupling forces the $d$-vector to align parallel or anti-parallel
to the AFM order parameter. It also promotes the emergence of two
collective modes in the AFM-SC coexistence state: the first one is
a Goldstone mode related to the precession of the $d$-vector around
the staggered magnetization. The second one is a massive mode that
is nearly degenerate with the Leggett-type mode associated with the
relative phase between the singlet and triplet components of the SC
order parameter. The experimental detection of these modes would provide
unambiguous evidence for a microscopic AFM-SC state, in contrast to
the more trivial situation of phase separated domains displaying either
AFM or SC order, but not both.

We also went beyond the Ginzburg-Landau mean-field approach and studied
the impact of Gaussian magnetic fluctuations on the phase diagram.
We found that, as expected, the inclusion of these fluctuations acts
mainly to shrink the region where AFM order exists, while at the same
time keeping the second-order nature of the phase transition lines
and tetracritical character of the multi-critical point. Our main
result is that, despite the fact that AFM and SC are competing orders,
the coupling between magnetic and t-SC degrees of freedom always favors
an enhancement of the AFM-SC coexistence state. Although in this paper
we considered a particular two-band microscopic model, which has been
widely employed in the study of iron-based superconductors, much of
our conclusions rely solely on the properties of the Ginzburg-Landau
free energy, such as the AFM-singlet SC attraction promoted by the
t-SC degrees of freedom, the coupling between the triplet $d$-vector
and the staggered magnetization, and the nature of the collective
modes inside the AFM-SC coexistence state. Consequently, we expect
these results to be relevant not only for iron pnictides, but also
for cuprates and heavy fermions. Overall, the impact of the t-SC degrees
of freedom on the phase diagram of competing AFM and SC states highlights
the importance of composite orders arising in the regime where two
distinct types of order have comparable energies, illustrating that
their interplay goes beyond just the competition for the same electronic
states. 

\subsection*{Acknowledgments}

We thank D. Agterberg, A. Chubukov, J. Kang, and M. Schuett for useful
discussions. DEA was supported by the Brazilian agencies CNPq and
CAPES Foundation, Ministry of Education of Brazil, Brasília - DF 70.040-020,
Brazil. RMF is supported by the U.S. Department of Energy, Office
of Science, Basic Energy Sciences, under Award number DE-SC0012336.
EM is supported by CNPq (grant number 304311/2010-3).

\appendix

\section{Ginzburg-Landau functional coefficients\label{sec:GL-coefficients}}

The coefficients of the GL expansion of the free energy density $f$
{[}Eq. (\ref{action-expanded-final}){]} are given by

\begin{equation}
a_{\mathrm{m}}=\frac{4}{V_{\mathrm{m}}}+4\int_{k}G_{1,k}G_{2,k}\mathrm{,}\quad u_{\mathrm{m}}=4\int_{k}G_{1,k}^{2}G_{2,k}^{2}\mathrm{,}\label{eq:am_um}
\end{equation}
\begin{equation}
a_{\mathrm{s}}=\frac{4}{V_{\mathrm{s}}}-2\sum_{\alpha}\int_{k}G_{\alpha,k}G_{\alpha,-k}\mathrm{,}\quad u_{\mathrm{s}}=2\sum_{\alpha}\int_{k}G_{\alpha,k}^{2}G_{\alpha,-k}^{2}\mathrm{,}\label{eq:as_us}
\end{equation}
\begin{equation}
a_{\mathrm{t}}=\frac{4}{V_{\mathrm{t}}}-4\int_{k}G_{1,k}G_{2,-k}\mathrm{,}\quad u_{\mathrm{t}}=4\int_{k}G_{1,k}^{2}G_{2,-k}^{2}\mathrm{,}\label{eq:at_ut}
\end{equation}
\begin{equation}
\lambda=-2\sum_{\alpha}\int_{k}G_{\alpha,k}G_{\alpha,-k}(G_{\bar{\alpha},k}+G_{\bar{\alpha},-k})\mathrm{,}\label{eq:lambda}
\end{equation}
\begin{equation}
\gamma_{12}=4\int_{k}G_{1,k}G_{1,-k}G_{2,k}G_{2,-k}\mathrm{,}\label{eq:gamma12}
\end{equation}
\begin{equation}
\gamma_{\mathrm{ms}}=-4\sum_{\alpha}\int_{k}G_{\alpha,k}^{2}G_{\alpha,-k}G_{\bar{\alpha},k}-\gamma_{12}\mathrm{,}\label{eq:gamma_ms}
\end{equation}
\begin{equation}
\gamma_{\mathrm{mt}}=-4\sum_{\alpha}\int_{k}G_{\alpha,k}^{2}G_{\bar{\alpha},k}G_{\bar{\alpha},-k}-\gamma_{12}\label{eq:gamma_mt}
\end{equation}
and
\begin{equation}
\gamma_{\mathrm{st}}=4\sum_{\alpha}\int_{k}G_{\alpha,k}^{2}G_{\alpha,-k}G_{\bar{\alpha},-k}+\gamma_{12}\mathrm{,}\label{eq:gamma_st}
\end{equation}
where $G_{1,k}^{-1}=i\omega_{n}-\xi_{1,\boldsymbol{k}}$ and $G_{2,k}^{-1}=i\omega_{n}-\xi_{2,\boldsymbol{k}+\boldsymbol{Q}}$
are the non-interacting Green's functions for each band. Note that
the symmetry $\xi_{2,\boldsymbol{k}+\boldsymbol{Q}}=\xi_{2,\boldsymbol{k}-\boldsymbol{Q}}$
implies that $G_{2,-k}^{-1}=-i\omega_{n}-\xi_{2,-\boldsymbol{k}+\boldsymbol{Q}}=-i\omega_{n}-\xi_{2,-\boldsymbol{k}-\boldsymbol{Q}}$. 

In order to gain more analytical insight, we have made the following
simplifications, following Ref. \cite{Chubukov09}: $\delta_{0}\left(k\right)\approx\delta_{0}\left(k_{F}\right)\equiv\delta_{0}$
and $\delta_{2}\left(k\right)\approx\delta_{2}\left(k_{\mathrm{F}}\right)\equiv\delta_{2}$,
so that $\delta_{\boldsymbol{k}}\to\delta_{\theta}=\delta_{0}+\delta_{2}\cos(2\theta)$.
Here, $k_{\mathrm{F}}$ is the Fermi wave vector, defined so that
$\xi_{1,k_{\mathrm{F}}}=0$ and $k_{\mathrm{F}}^{2}/2m=\varepsilon_{1,0}-\mu\equiv\xi_{\mathrm{F}}$.
Thus, we can write the dispersions as $\xi_{1,\boldsymbol{k}}=\xi_{k}=\xi_{\mathrm{F}}-\frac{k^{2}}{2m}$
and $\xi_{2,\boldsymbol{k}+\boldsymbol{Q}}=-\xi_{k}+2\delta_{\theta}$.
This allows us to write $\frac{1}{\upsilon}\sum_{\boldsymbol{k}}\to m\int_{0}^{2\pi}\frac{d\theta}{2\pi}\int_{-\infty}^{\xi_{\mathrm{F}}}\frac{d\xi}{2\pi}$;
since we consider the case $\xi_{F}\ll T$, we can generally send
$\xi_{\mathrm{F}}\to\infty$ in the upper limit of the integral, provided
that the integrand does not vanish. Notice that $m$ is proportional
to the two-dimensional density of states.

Carrying out the integrations over momentum and frequency we obtain

\begin{equation}
a_{\mathrm{s}}=\frac{2m}{\pi}\log(T/T_{\mathrm{c},0})\mathrm{,}\quad u_{\mathrm{s}}=\frac{7\zeta(3)m}{4\pi^{3}T^{2}}\mathrm{,}\label{eq:as_us-final}
\end{equation}
where $T_{\mathrm{c},0}=(2\xi_{\mathrm{F}}/\pi)e^{\gamma-2\pi/mV_{\mathrm{s}}}$
and $\gamma\approx0.577$ is Euler-Mascheroni's constant. Note that
these couplings do not depend on the parameters $\delta_{0}$ and
$\delta_{2}$. Furthermore, $a_{\mathrm{t}}=4/V_{\mathrm{t}}-2m/\pi$
and $u_{\mathrm{t}}=0$ (in the limit of $\xi_{\mathrm{F}}/T\to\infty$),
and 
\begin{equation}
a_{\mathrm{m}}=\frac{2m}{\pi}\log(T/\bar{T}_{\mathrm{N},0})+\frac{2m}{\pi}\tilde{a}_{\mathrm{m}}\left(\tilde{\delta}_{0},\tilde{\delta}_{2}\right)\mathrm{,}\label{eq:am}
\end{equation}
where $\bar{T}_{\mathrm{N},0}=(2\xi_{\mathrm{F}}/\pi)e^{\gamma-2\pi/mV_{\mathrm{m}}}$,
$\tilde{\delta}_{0\,(2)}=\delta_{0\,(2)}/2\pi T$, and 
\begin{eqnarray}
\tilde{a}_{\mathrm{m}}\left(\tilde{\delta}_{0},\tilde{\delta}_{2}\right) & = & \gamma+\log4\nonumber \\
 &  & +\frac{1}{2}\left\langle \psi^{(0)}\left(\frac{1}{2}+\mathrm{i}\tilde{\delta}_{\theta}\right)+\psi^{(0)}\left(\frac{1}{2}-\mathrm{i}\tilde{\delta}_{\theta}\right)\right\rangle _{\theta}\mathrm{,}\label{eq:am_tilde}
\end{eqnarray}
where $\tilde{\delta}_{\theta}=\delta_{\theta}/2\pi T$, $\psi^{(0)}(z)$
is the digamma function and the angular brackets denote angular averages
$\left\langle \cdots\right\rangle _{\theta}=\int_{0}^{2\pi}\frac{d\theta}{2\pi}(\cdots)$.

Finally, introducing the dimensionless quantities $\tilde{u}_{\mathrm{m}}=4\pi^{3}T^{2}u_{\mathrm{m}}/m$,
$\tilde{\lambda}=\pi^{2}T\lambda/m$, $\tilde{\gamma}_{\mathrm{st}}=2\pi^{3}T^{2}\gamma_{\mathrm{st}}/m$,
$\tilde{\gamma}_{\mathrm{ms}}=4\pi^{3}T^{2}\gamma_{\mathrm{ms}}/m$
and $\tilde{\gamma}_{\mathrm{mt}}=-2\pi^{3}T^{2}\gamma_{\mathrm{mt}}/m$
we find
\begin{equation}
\tilde{u}_{\mathrm{m}}=-\frac{1}{4}\left\langle \psi^{(2)}\left(\frac{1}{2}+\mathrm{i}\tilde{\delta}_{\theta}\right)+\psi^{(2)}\left(\frac{1}{2}-\mathrm{i}\tilde{\delta}_{\theta}\right)\right\rangle _{\theta}\mathrm{,}\label{eq:um_tilde}
\end{equation}
\begin{equation}
\tilde{\lambda}=\left\langle \frac{\gamma+\log4}{\tilde{\delta}_{\theta}}+\frac{\psi^{(0)}\left(\frac{1}{2}+\mathrm{i}\tilde{\delta}_{\theta}\right)+\psi^{(0)}\left(\frac{1}{2}-\mathrm{i}\tilde{\delta}_{\theta}\right)}{2\tilde{\delta}_{\theta}}\right\rangle _{\theta}\mathrm{,}\label{eq:lambda_tilde}
\end{equation}
\begin{equation}
\tilde{\gamma}_{\mathrm{st}}=\left\langle \frac{\gamma+\log4}{\tilde{\delta}_{\theta}^{2}}+\frac{\psi^{(0)}\left(\frac{1}{2}+\mathrm{i}\tilde{\delta}_{\theta}\right)+\psi^{(0)}\left(\frac{1}{2}-\mathrm{i}\tilde{\delta}_{\theta}\right)}{2\tilde{\delta}_{\theta}^{2}}\right\rangle _{\theta}\mathrm{,}\label{eq:gamma_st_tilde}
\end{equation}
\begin{equation}
\tilde{\gamma}_{\mathrm{mt}}=\tilde{\gamma}_{\mathrm{st}}-\mathrm{i}\left\langle \frac{\psi^{(1)}\left(\frac{1}{2}+\mathrm{i}\tilde{\delta}_{\theta}\right)-\psi^{(1)}\left(\frac{1}{2}-\mathrm{i}\tilde{\delta}_{\theta}\right)}{4\tilde{\delta}_{\theta}}\right\rangle _{\theta}\label{eq:gamma_mt_tilde}
\end{equation}
and $\tilde{\gamma}_{\mathrm{ms}}=\tilde{\gamma}_{\mathrm{st}}-2\tilde{\gamma}_{\mathrm{mt}}$.
Moreover, $\gamma_{\mathrm{st}}=2\gamma_{12}>0$. In the expressions
above, $\psi^{(n)}(z)$ is the polygamma function of order $n$, defined
as $\psi^{(n)}(z)=\frac{d^{n+1}}{dz^{n+1}}\log[\Gamma(z)]$, where
$\Gamma(z)$ is the gamma function. Away from the perfect nesting
condition we can only evaluate these angular averages numerically.

When the staggered magnetization is not a homogeneous function of
space and (imaginary) time, the quadratic term in $\boldsymbol{M}$
in the GL expansion {[}Eq.~(\ref{action-expanded-final}){]} becomes
$\frac{1}{2}\int_{q}\chi_{\mathrm{m}}^{-1}(\boldsymbol{q},\nu_{n})M_{q}^{2}$,
where 
\begin{equation}
\chi_{\mathrm{m}}^{-1}(\boldsymbol{q},\nu_{n})=\frac{4}{V_{\mathrm{m}}}+4\int_{k}G_{2,k}G_{1,k-q}\label{eq:chi_m}
\end{equation}
is the frequency- and momentum-dependent magnetic susceptibility.
In the static limit, $\nu_{n}=0$, and for small $\boldsymbol{q}=|\boldsymbol{q}|(\cos\theta_{q},\sin\theta_{q})$,
this quantity naturally has the same anisotropy as the Fermi surface
\begin{equation}
\chi_{\mathrm{m}}^{-1}(|\boldsymbol{q}|\ll1,\nu_{n}=0)\equiv\chi_{\boldsymbol{q}}^{-1}=a_{\mathrm{m}}+q^{2}(\kappa+\kappa{}_{2}\cos2\theta_{q}),\label{eq:chi_final}
\end{equation}
where\begin{widetext} 
\begin{equation}
\kappa=\frac{1}{64\pi^{2}T}\left\langle \left(\tilde{\delta}_{\theta}-\frac{\xi_{\mathrm{F}}}{2\pi T}\right)\left[\psi^{(2)}\left(\frac{1}{2}+\mathrm{i}\tilde{\delta}_{\theta}\right)+\psi^{(2)}\left(\frac{1}{2}-\mathrm{i}\tilde{\delta}_{\theta}\right)\right]\right\rangle _{\theta}\label{eq:kappa}
\end{equation}
and
\begin{eqnarray}
\kappa_{2} & = & \frac{1}{64\pi^{2}T}\left\langle \cos2\theta\left(\tilde{\delta}_{\theta}-\frac{\xi_{\mathrm{F}}}{2\pi T}\right)\left[\psi^{(2)}\left(\frac{1}{2}+\mathrm{i}\tilde{\delta}_{\theta}\right)+\psi^{(2)}\left(\frac{1}{2}-\mathrm{i}\tilde{\delta}_{\theta}\right)\right]\right\rangle _{\theta}\nonumber \\
 &  & +\frac{\mathrm{i}}{32\pi^{2}T}\left\langle \cos2\theta\left[\psi^{(1)}\left(\frac{1}{2}+\mathrm{i}\tilde{\delta}_{\theta}\right)-\psi^{(1)}\left(\frac{1}{2}-\mathrm{i}\tilde{\delta}_{\theta}\right)\right]\right\rangle _{\theta}\mathrm{.}\label{eq:kappa_2}
\end{eqnarray}
\end{widetext}Note that $\kappa_{2}=0$ when $\delta_{2}=0$, i.e.,
when $\delta_{\theta}$ does not depend on the angle $\theta$. For
simplicity, we will neglect the anisotropy so that we can write $\chi_{\boldsymbol{q}}^{-1}=a_{\mathrm{m}}+\kappa q^{2}$.

\section{Minimization with respect to $\alpha_{\mathrm{m}d}$ and $\alpha_{\mathrm{st}}$
\label{sec:Minimization}}

In addition to the solutions $\sin\alpha_{\mathrm{st}\,0}=\sin\alpha_{\mathrm{m}d\,0}=0$,
the extremum conditions $\partial_{\alpha_{\mathrm{st\,0}}}f=0$ and
$\partial_{\alpha_{\mathrm{m}d\,0}}f=0$ in Eqs.~(\eqref{df/dast})
and (\eqref{df/damd}) can also be satisfied by Eqs.~(\eqref{alphast0alt})
and (\eqref{alphamd0alt}), respectively. There are, therefore, four
combinations that solve Eqs.~(\eqref{df/dast}) and (\eqref{df/damd}):
(i) $\sin\alpha_{\mathrm{st}\,0}=\sin\alpha_{\mathrm{m}d\,0}=0$,
(ii) Eqs.~(\eqref{alphast0alt}) and (\eqref{alphamd0alt}), (iii)
$\sin\alpha_{\mathrm{st}\,0}=0$ and Eq.~(\eqref{alphamd0alt}),
and (iv) and Eq.~(\eqref{alphast0alt}) and $\sin\alpha_{\mathrm{m}d\,0}=0$.

Case (i) was studied in the main text and the phase diagram was presented
in Figure~\ref{fig:phase-diagram}.

The equations of case (ii) require, for consistency, that $4\gamma_{12}^{2}|\Delta_{\mathrm{t}\,0}|^{2}\cos\alpha_{\mathrm{st}\,0}=-\lambda^{2}\cos\alpha_{\mathrm{st}\,0}$
as well as $4\gamma_{12}^{2}|\Delta_{\mathrm{t}\,0}|^{2}\cos\alpha_{\mathrm{m}d\,0}=-\lambda^{2}\cos\alpha_{\mathrm{m}d\,0}$.
These equations, on the other hand, can only be satisfied if $\cos\alpha_{\mathrm{st}\,0}=\cos\alpha_{\mathrm{m}d\,0}=0$,
because $\lambda^{2}\neq-4\gamma_{12}^{2}|\Delta_{\mathrm{t}\,0}|^{2}$.
In the model we are considering, $\gamma_{\mathrm{st}}=2\gamma_{12}$.
Therefore, $\partial_{|\Delta_{\mathrm{t}\,0}|}F=0$ leads to $M_{0}^{2}=-a_{\mathrm{t}}/(\gamma_{\mathrm{mt}}+\gamma_{\mathrm{st}})$,
if $|\Delta_{\mathrm{t}}|\neq0$. We know that $a_{\mathrm{t}}>0$
and we found numerically that $\gamma_{\mathrm{mt}}+\gamma_{\mathrm{st}}>0$.
Since $M_{0}^{2}$ cannot be negative, this is not a physical solution.

For case (iii) we have 
\begin{equation}
\frac{\partial^{2}f}{\partial\alpha_{\mathrm{st}\,0}^{2}}=-2\gamma_{12}|\Delta_{\mathrm{s}\,0}|^{2}|\Delta_{\mathrm{t}\,0}|^{2}-\frac{\lambda^{2}|\Delta_{\mathrm{s}\,0}|^{2}}{2\gamma_{12}},
\end{equation}
and $\frac{\partial^{2}f}{\partial\alpha_{\mathrm{st}\,0}\partial\phi_{0}^{i}}=0$
for $\phi_{0}^{i}\neq\alpha_{\mathrm{st}}$. Thus, $\frac{\partial^{2}f}{\partial\alpha_{\mathrm{st}\,0}^{2}}$
is an eigenvalue of the Hessian matrix $(\mathbb{H})_{\{\phi_{0}^{i}\}}$.
In our microscopic model $\gamma_{12}$ is strictly positive and thus
$\partial^{2}f/\partial\alpha_{\mathrm{st}\,0}^{2}<0$. We conclude
that case (iii) does not correspond to a local minimum of the free
energy.

Finally, in case (iv) we obtain 
\begin{equation}
\partial_{M}f=(a_{\mathrm{m}}+u_{\mathrm{m}}M^{2}+\gamma_{\mathrm{ms}}|\Delta_{\mathrm{s}}|^{2}+\gamma_{\mathrm{mt}}|\Delta_{\mathrm{t}}|^{2}-\lambda^{2}/2\gamma_{12})M,
\end{equation}
and the conditions $\partial_{|\Delta_{\mathrm{s}}|}f=0$ and $\partial_{|\Delta_{\mathrm{t}}|}f=0$
lead to, respectively 
\begin{eqnarray}
[a_{\mathrm{s}}+u_{\mathrm{s}}|\Delta_{\mathrm{s}}|^{2}+\gamma_{\mathrm{ms}}M^{2}+(\gamma_{\mathrm{st}}-2\gamma_{12})|\Delta_{\mathrm{t}}|^{2}]|\Delta_{\mathrm{s}}|^{2} & = & 0,\\{}
[a_{\mathrm{t}}+\gamma_{\mathrm{mt}}M^{2}+(\gamma_{\mathrm{st}}-2\gamma_{12})|\Delta_{\mathrm{s}}|^{2}]|\Delta_{\mathrm{t}}|^{2} & = & 0.
\end{eqnarray}
Therefore, for our model (with $\gamma_{\mathrm{st}}=2\gamma_{12}$)
we get $M_{0}^{2}=-a_{\mathrm{t}}/\gamma_{\mathrm{mt}}$ (if $|\Delta_{\mathrm{t}}|^{2}\neq0$),
and 
\begin{eqnarray}
|\Delta_{\mathrm{t}\,0}|^{2} & = & (\lambda^{2}/\gamma_{\mathrm{st}}-a_{\mathrm{m}}-u_{\mathrm{m}}M_{0}^{2}-\gamma_{\mathrm{ms}}|\Delta_{\mathrm{s}\,0}|^{2})/\gamma_{\mathrm{mt}},\\
|\Delta_{\mathrm{s}\,0}|^{2} & = & -(a_{\mathrm{s}}+M_{0}^{2}\gamma_{\mathrm{ms}})/u_{\mathrm{s}}
\end{eqnarray}
For the two-band model $M_{0}^{2}=-a_{\mathrm{t}}/\gamma_{\mathrm{mt}}$
is always positive because $\gamma_{\mathrm{mt}}<0$. We have computed
the eigenvalues of the Hessian matrix in the regions where both $|\Delta_{\mathrm{s}\,0}|^{2}$
and $|\Delta_{\mathrm{t}\,0}|^{2}$ are positive and found that there
is at least one negative eigenvalue. This means that the free energy
is not a minimum at this solution. Thus, we conclude that, at least
for the two-band model studied here, $\hat{\boldsymbol{M}}\cdot\hat{\boldsymbol{d}}=\pm1$
and $\alpha_{\mathrm{st}}=0$ or $\alpha_{\mathrm{st}}=\pi$.

\end{document}